\documentclass[%
 reprint,
 amsmath,amssymb,
prb,
nofootinbib,
floatfix,superscriptaddress,longbibliography
]{revtex4-2}
\usepackage{empheq}
\usepackage{bbm}
\usepackage{float}
\usepackage[normalem]{ulem}
\usepackage{lipsum}
\usepackage{makecell}
\usepackage{mathtools, cuted, relsize}
\usepackage{esvect}
\usepackage{soul, xcolor}
\usepackage[caption = false]{subfig}
\usepackage{graphicx}
\usepackage{dcolumn,multirow, tabularx}
\usepackage{bm}
\usepackage{hyperref}
\usepackage{float}
\usepackage{physics}
\usepackage{slashed}
\usepackage{balance}
\usepackage[mathlines]{lineno}
\usepackage{mathtools}
\mathtoolsset{showmanualtags}
\RequirePackage{fix-cm}

\begin{document}

\title{Breaking and resurgence of symmetry in the non-Hermitian Su-Schrieffer-Heeger model in photonic waveguides}

\author{E. Slootman}
\thanks{Authors contributed equally}
\affiliation{Institute for Theoretical Physics, Utrecht University, Princetonplein 5, 3584CC Utrecht, The Netherlands \looseness=-1}%
\affiliation{Adaptive Quantum Optics (AQO), MESA\textsuperscript{+} Institute for Nanotechnology, University of Twente, PO Box 217, 7500 AE Enschede, The Netherlands}
\author{W. Cherifi}
\thanks{Authors contributed equally}
\affiliation{Department of Physics, Stockholm University, S-10691, Stockholm, Sweden}
\author{L. Eek}
 \affiliation{Institute for Theoretical Physics, Utrecht University, Princetonplein 5, 3584CC Utrecht, The Netherlands \looseness=-1}%
\author{R. Arouca}
\affiliation{Department of Physics and Astronomy, Uppsala University, Uppsala, Sweden}
\author{E. J. Bergholtz}
\affiliation{Department of Physics, Stockholm University, S-10691, Stockholm, Sweden}
\author{M. Bourennane}
\thanks{Corresponding author: boure@fysik.su.se}
\affiliation{Department of Physics, Stockholm University, S-10691, Stockholm, Sweden}
\author{C. Morais Smith}
\affiliation{Institute for Theoretical Physics, Utrecht University, Princetonplein 5, 3584CC Utrecht, The Netherlands \looseness=-1}%

\date{\today}

\begin{abstract}
Symmetry is one of the cornerstones of modern physics and has profound implications in different areas. In symmetry-protected topological systems, symmetries are responsible for protecting surface states, which are at the heart of the fascinating properties exhibited by these materials. When the symmetry protecting the edge mode is broken, the topological phase becomes trivial. By engineering losses that break the symmetry protecting a topological Hermitian phase, we show that a new genuinely non-Hermitian symmetry emerges, which protects and selects one of the boundary modes: the topological monomode. Moreover, the topology of the non-Hermitian system can be characterized by an effective Hermitian Hamiltonian in a higher dimension. To corroborate the theory, we experimentally investigated the non-Hermitian 1D and 2D SSH models using photonic lattices and observed dynamically generated monomodes in both cases. We classify the systems in terms of the (non-Hermitian) symmetries that are present and calculate the corresponding topological invariants. 

\end{abstract}

\maketitle

\section{Introduction}
Ever since the first observation of topological behaviour through the quantum Hall effect, topological states of matter have allowed for an entirely new perspective on condensed-matter physics \cite{Hasan2010, Bernevig2013, Qi2011}. These states challenge the Ginzburg-Landau classification of phases of matter in terms of spontaneous symmetry breaking, being instead characterized by an underlying topology. The topological robustness of these materials is manifest in both the presence of topological invariants at their bulk and topologically protected states at their boundaries. Because of their dissipationless nature, topological boundary modes are expected to become a key ingredient in nanotechnology. Moreover, the zero-energy topological corner states that may appear at the edges of 1D and 2D systems are promising candidates to realize qubits in the field of quantum computing and spintronics \cite{Pesin2012, Sarma2015, He2019}. For special classes of these materials, the corner states are related by symmetry, which makes them come in pairs. A paradigmatic example is the Su-Schrieffer-Heeger (SSH) model \cite{Su1979}, which exhibits two edge states related by sublattice and inversion symmetry. Because these two modes are connected and located at different ends of the lattice, they are jointly protected against symmetry-preserving perturbations. The same argument also applies to higher-order topology, where multiple symmetry-related boundary modes appear at corners (2D) or hinges (3D) \cite{Benalcazar2017a}. An intriguing unanswered question is whether these topological pairs could be broken, and a monomode located at one single edge would still be topologically stable.

Recently, the understanding of topological states was greatly enlarged by non-Hermitian Hamiltonians \cite{Gong2018,Kawabata2019, Bergholtz2021, Zeuner2015, Pan2018, Helbig2020, Xiao2020, Ghatak2020}, in many ways. Unique non-Hermitian topological phenomena were revealed, like the non-Abelian topological properties of exceptional points \cite{Miri2019, Berry2004}, or the non-Hermitian skin effect protected by a spectral winding number \cite{Bergholtz2021, Lee2016, Kunst2018, Yao2018, Esaki2011, Herviou2019, Okuma2023, Lin2023}. By allowing the Hamiltonian to be non-Hermitian, one extends the different symmetry-protected phases \cite{Gong2018, Kawabata2019}. Particularly interesting is to investigate how structured loss, which acts as dissipation and would naively be expected to be detrimental to topology, can actually be harnessed for robustly targeting boundary modes, as shown in Ref. \cite{Cherifi2023}.
\begin{figure*}
    \centering
    \includegraphics[]{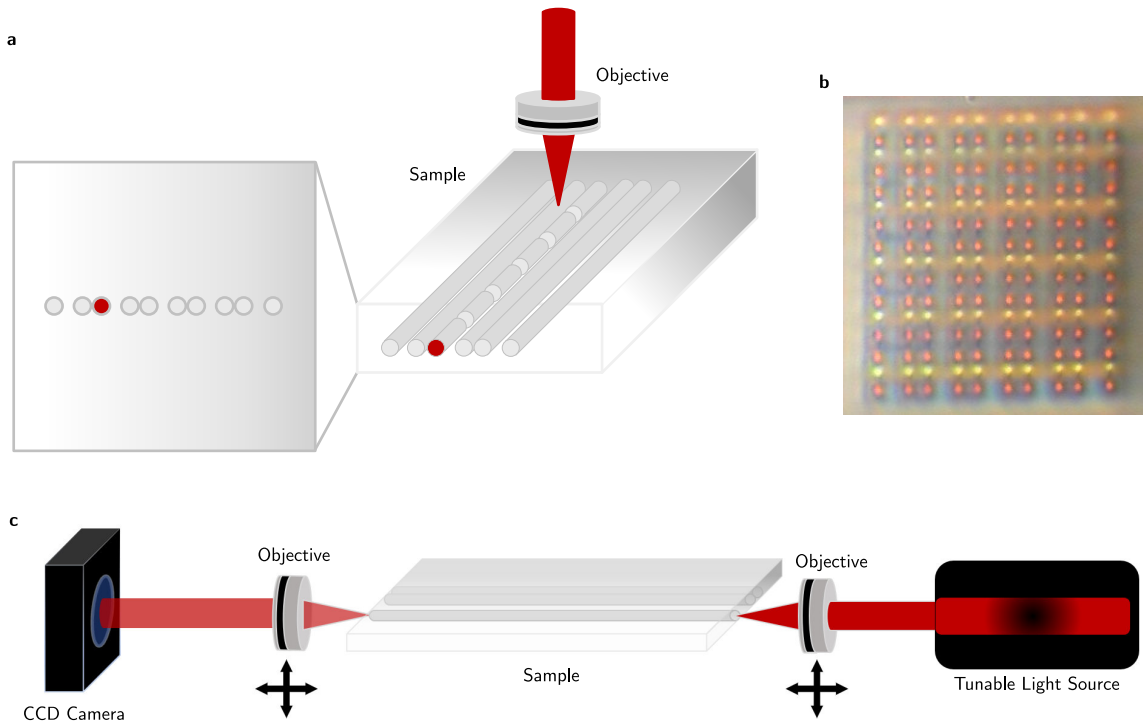}
    \caption{Overview of the experimental setup. \textbf{a}, Schematic overview of the waveguides fabrication. The loss is introduced by making cuts in the waveguide. The light is focused using an objective. \textbf{b}, A microscopic picture of a produced 2D SSH sample. Here, $t_1= 11\ \mu m$ and $t_2= 7\ \mu m$. \textbf{c}, Schematic overview of the measurement process. The light from the tunable light source is focused using an objective into a specific waveguide of the sample. At the other end, a camera captures the resulting light.}
    \label{fig:experimentalsetup}
\end{figure*}

Here, we explore the consequences of non-Hermitian phenomena to design and experimentally observe how Hermitian symmetries morph into non-Hermitian ones. Specifically, we consider how losses, which explicitly break a symmetry of the system, affect symmetry-related boundary modes, while retaining a generalized non-Hermitian symmetry constraint. We consider loss localized in one sublattice of non-Hermitian generalizations of the SSH model in 1D and 2D, and probe the effects of both, boundaries and topological defects. The addition of loss to one sublattice \textit{breaks} the (Hermitian) sublattice symmetry, and one would expect that these edge states would be destroyed. Nevertheless, we found that the two edge modes \textit{remain}, and one of them decays in time. We established that this apparent contradiction can be understood by considering \textit{how Hermitian sublattice symmetry turns into a non-Hermitian sublattice symmetry} (which is preserved by this loss configuration), and how the topology of this \textit{non-Hermitian} system can be characterized by an \textit{effective Hermitian} Hamiltonian in a higher dimension. The same holds for the 2D SSH model. To corroborate our theory, we engineer structured loss in a system of coupled waveguides, see Fig.~\ref{fig:experimentalsetup}. A schematic overview of the optical waveguide writing process is shown in Fig.~\ref{fig:experimentalsetup}a. Fig.~\ref{fig:experimentalsetup}b illustrates a 2D SSH sample. Finally, Fig.~\ref{fig:experimentalsetup}c gives a schematic overview of the measurement setup. Details of the experimental setup and sample preparation are given in Sec.~\ref{sec:exp}. 

The propagation across the waveguides is described by a Schrödinger-like equation \cite{Longhi2009}, such that the propagation across lossy waveguides can be represented approximately by non-Hermitian Hamiltonians \cite{Weimann2016}. Even for the simple case of loss in only one waveguide, we find that some of these corner zero-modes acquire a finite imaginary part, while others remain pinned at zero energy. Due to these spectral properties, just a subset of the topological modes survives in time, characterizing isolated monomodes in 1D and 2D. However, all the boundary modes are still related by symmetry, which allows us to show that these modes are robust and characterized by topological invariants.
\begin{figure*}
    \centering
    \includegraphics{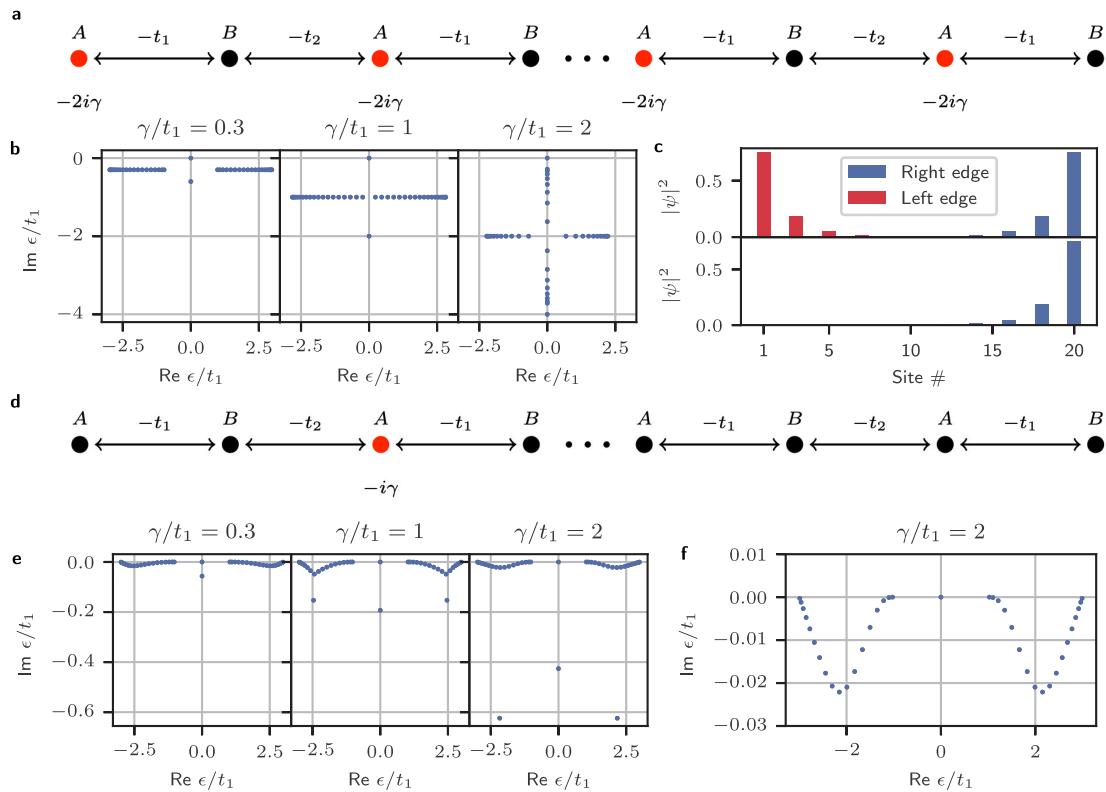}
    \caption{Theoretical description of an SSH chain with loss. \textbf{a}, 1D SSH lattice, described by Eq.~\eqref{eq:hamiltonian}. \textbf{b}, Spectrum of the SSH model with alternating loss, for OBC in the topological regime, with $t_2/t_1=2$. The system consists of 40 lattice sites. The energies are shown in the complex plane, for different values of $\gamma/t_1$. \textbf{c}, Time evolution of the eigenstates of the Hamiltonian given by Eq.~\eqref{eq:hamiltonian}. The two edge modes, for $t_2/t_1=2$ and $\gamma/t_1=2$, are depicted at $t=0$ (top) and at $t=10$ (bottom) for 20 lattice sites. One observes that after some time, the left edge mode has disappeared, giving rise to a monomode. The time is measured in units of ($\hbar/t_1$) \textbf{d}, 1D SSH lattice with only one lossy site. \textbf{e}, Spectrum of the SSH model with only one lossy site, for OBC in the topological regime, with $t_2/t_1=2$. The system consists of 40 lattice sites. The energies are shown in the complex plane, for different values of $\gamma/t_1$. \textbf{f}, Zoom in on the last panel of \textbf{e}. Part of the spectrum is cropped out, to show the behaviour of the imaginary part of the bulk.}
    \label{fig:chainspec}
\end{figure*}

The dynamical preparation of the monomodes set our work apart from other realizations of single isolated topological modes, like the odd site SSH chain \cite{Ghatak2020}, defective non-Hermitian SSH boundary states \cite{Lee2016, Kunst2018, Yao2018}, band structure monopoles \cite{Pap2018, Konig2022}, or defect states \cite{Weimann2016, Poli2015, Noh2018}. By exploring injection of light in multiple sites, we also show that this differs from previous works that showed isolated modes due to the initial state preparation \cite{ElHassan2019, Cerjan2020}. In contrast, we present a simple and generic way to prepare the isolated boundary modes by engineering loss, while preserving the topological aspects of these states, as shown by explicit calculation of the topological invariants. The understanding of how topological states may remain robust despite breaking the symmetry that protects them brings a new perspective into the field of non-Hermitian topological insulators. 

The outline of this paper is the following: In Sec.~\ref{sec:model}, we present the SSH model with loss that we investigate in this work, and show the emergence of a monomode. In Sec.~\ref{sec:exp1} a comparison between theory and experiment is provided for 1D and 2D systems. Finally, the topological properties are analyzed using Chern numbers and the Jackiw-Rebbi method in Sec.~\ref{sec:topo}. Our conclusions are presented in Sec.~\ref{sec:conc}. The appendices list additional details of the experimental and theoretical procedure for completeness.

\section{SSH models with loss}\label{sec:model}
We consider a version of the SSH model, in which we introduce a loss of strength $2\gamma$ on the $A$ sublattice \cite{Miri2019, Weimann2016, Dangel2018a, Roccati2022}, see Fig.~\ref{fig:chainspec}a. This model and others with local gain and loss are discussed in more detail in Appendix \ref{app:gainandloss}, see also Figs.~\ref{fig:gainandlosschain}-\ref{fig:s10} \footnote{An additional model of losses over even sites of a Hermitian SSH where an edge state is present together with the appearance of bulk skin effect is presented in Ref. ~\cite{martinez-strasser}.}. The Hamiltonian of this system is given by 
\begin{equation}
\hspace{-0.2cm} H =  -\sum_{n=1}^N \Big[ \left(t_1 a_n^\dagger b_n+ t_2b_n^\dagger  a_{n+1}+H.c.\right) + 2i \gamma a_n^\dagger a^{}_n \Big],
    \label{eq:hamiltonian}
\end{equation}
where $N$ is the number of unit cells, and $t_1, t_2$ denote, respectively, the inter- and the intra-cell hopping parameter. For simplicity, we assume them to be real. The operators $a_n$ ($a_n^\dagger$) annihilate (create) a particle in sublattice $A$ at site $n$ (similarly for $B$). The corresponding dispersion relation for periodic boundary conditions (PBC) is given by
\begin{equation}
    \epsilon(k) = -i\gamma + \sqrt{-\gamma^2 + t_1^2 + t_2^2 + 2 t_1 t_2 \cos(ka)},
    \label{eq:spectrum}
\end{equation}
where the lattice parameter is denoted by $a$. We note that the spectrum of the system is similar to the one with open boundary conditions (OBC). This spectrum is also like the one of a chain with alternating gain and loss, shifted down on the imaginary axis by $\gamma$ (see Appendix \ref{app:gainandloss}). The Hermitian SSH model ($\gamma=0$) has a topological phase for $|t_2|>|t_1|$. In the case of OBC, this manifests itself in a pair of zero-energy edge modes. Surprisingly, when $\gamma$ is taken to be finite, the edge modes persist, even if the symmetry protecting them has been broken. In addition, one of the zero modes acquires a negative imaginary energy, while the other one remains at zero energy. This occurs because we have introduced loss only on one of the two sublattices. Each edge mode has support on one sublattice, thus exhibiting different energies. This behaviour is illustrated in Fig.~\ref{fig:chainspec}b.

The Hamiltonian directly encodes the information about the time-evolution of the eigenstates through 
\begin{equation}
    | \psi(t) \rangle = U | \psi(0) \rangle = \text{e}^{-\frac{i}{\hbar} H t} | \psi(0) \rangle .
    \label{eq:time}
\end{equation}
Therefore, the amplitude of states with a negative (positive) imaginary energy will decrease (increase) over time. Since only one mode of the system has a non-negative imaginary energy, see Fig.~\ref{fig:chainspec}b, all but one of the states decays over time. This is illustrated in Fig.~\ref{fig:chainspec}c, where the time-evolution operator is applied to the two edge states of the system. It is clear that only the right edge mode, the one which has support on the sublattice without loss, will endure. Therefore, this system reveals the existence of a topological monomode.

The monomode can nevertheless be realized in a more straightforward way. Instead of including losses in every unit cell, it is sufficient to insert loss only on a single site in the chain, as is depicted in Fig.~\ref{fig:chainspec}d. The difference is that it will take longer for the bulk and the corresponding edge mode to decay. In addition, this decay time becomes size dependent. A second, more practical, constraint is that the lossy site should be ‘relatively’ close to the decaying edge mode, for the decay to occur more rapidly. Since the left (right) edge mode has support on the $A$ ($B$) sublattice, inserting the loss on the $B$ ($A$) sublattice near the left (right) edge will give the right (left) edge mode a very long lifetime. In this case, it would be difficult to observe the dynamically generated monomode. On the other hand, if we insert the loss on a site on sublattice $A$ in the second unit cell, corresponding to the Hamiltonian 
\begin{equation}
    H =  -\sum_{n=1}^N \left[ t_1 a_n^\dagger b^{}_n + t_2 b_n^\dagger  a^{}_{n+1} + H.c. \right] - i\gamma a_2^\dagger a^{}_2,
    \label{eq:hamiltonian-one}
\end{equation}
the left edge state will decay relatively fast. Nevertheless, some bulk states may take some time to decay. For a sufficiently long time, the left edge state and the bulk states belonging to the $A$ sublattice will always decay. This also becomes evident by analyzing the spectrum in the complex plane, shown in Fig.~\ref{fig:chainspec}e. All the bulk modes have acquired a small negative imaginary energy, see Fig.~\ref{fig:chainspec}f for a zoom in. $\mathcal{PT}$-symmetry, therefore, does not play a role in generating the topological monomode, since the SSH model with only one lossy site is no longer (passive) $\mathcal{PT}$-symmetric, as shown by its spectrum. The monomode is thus more fundamental and not simply a consequence of $\mathcal{PT}$-symmetry.

The monomode concept is not only applicable to the SSH model, but also to other models, even ones that are not in the Altland-Zirnbauer classification \cite{Altland1997, Ryu2010}. We make this explicit by considering higher-order topological (HOT) \cite{Benalcazar2017a, Benalcazar2017, Schindler2018} models. Now, we focus on the 2D SSH model,
\begin{equation}
\hspace{-0.2cm} H =  -\sum_{\langle\mathbf{r},\mathbf{r}'\rangle} \Big[ t_1 a_{\mathbf{r}}^\dagger b^{}_\mathbf{r}+ t_2b_\mathbf{r}^\dagger  a^{}_{\mathbf{r}'}+H.c.\Big],
    \label{eq:hamiltonian}
\end{equation}
which is a HOT metal \cite{Cerjan2020, Benalcazar2020, Cerjan2022}. In the above equation, $\mathbf{r}$ and $\mathbf{r'}$ denote first-neighbouring unit cells in a square lattice. Although the 2D SSH model presents bulk states at zero energy, they do not hybridize with the robust symmetry-protected corner modes for large system sizes. The four corner modes are protected by $C_{4v}$ and chiral symmetry, such that we can use again a similar insight to add loss in some of the sublattices and design only two or one corner state, depending on the loss distribution. 
\begin{figure*}[t]
 	\centering
 	\includegraphics{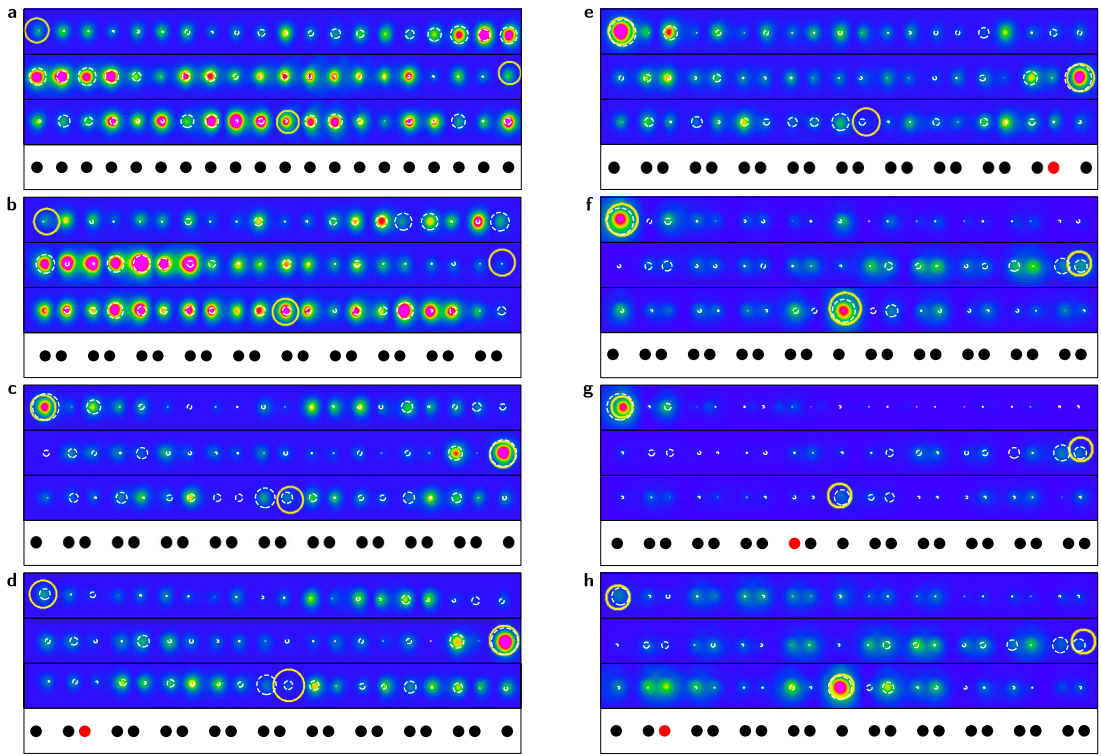}
 	\caption{Experimental realization of the monomode in 1D SSH chains. Yellow circles signal the point at which light was injected. The white dashed circles are the predictions of the tight-binding simulations. Their radius is proportional to the relative intensity. The lowest row of each subfigure is a schematic description of the lattice structure. Here, the red dot signals a lossy waveguide. The distance between waveguides has been exaggerated to allow for a prompt distinction between the trivial and topological regimes. \textbf{a}, Tight-binding chain (black dots) with $t_1=t_2$, dispersing into the bulk. \textbf{b}, Trivial SSH chain with $t_2/t_1=0.5$, dispersing into the bulk. \textbf{c}, Topological SSH chain without loss with $t_2/t_1=2$. Localised edge modes can be clearly observed. \textbf{d}, By adding loss ($\gamma/t_1=0.5$) at the red-dot site near the left edge, one of the edge modes disappears, revealing the monomode. \textbf{e}, When the loss is applied far away from the left edge, the corresponding edge mode does not decay within the experimental time scale. \textbf{f}, The inclusion of a topological defect into the system ($t_2/t_1=3.2$) does not affect the left edge mode, but leads to an additional mode pinned on the defect. \textbf{g}, The introduction of loss on the red-dot site near the defect destroys the mode at the defect and leads to a monomode at the edge. \textbf{h}, If instead the loss is engineered on the red-dot site near the left edge, the edge mode is destroyed, and a monomode occurs at the topological defect. For a separate visualization of numerical and experimental results, see Fig.~\ref{fig:S6} in the Appendix \ref{app:numpred}.}
 	\label{fig:exp} 	
 \end{figure*}

\begin{figure*}[t]
 	\centering
    \includegraphics{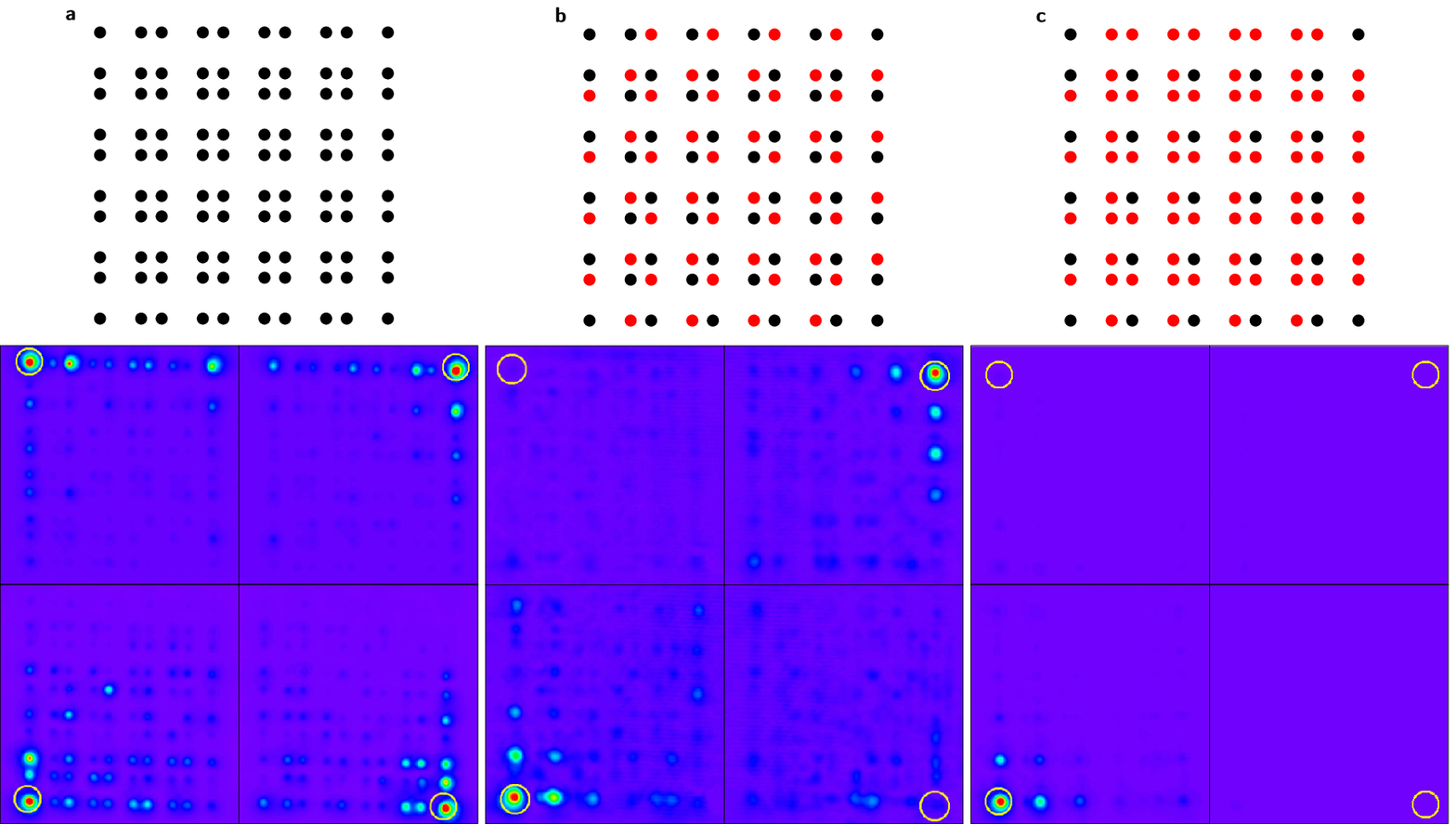}
    \caption{Experimental realization of bi- and monomodes in the 2D SSH model. Yellow circles signal the point at which light was injected. For numerical predictions, see Appendix \ref{app:numpred}. Top row shows the loss locations in the lattice (red dots) and the bottom row the experiments. Each quarter of the bottom images is one experimental realization, with light injected into the indicated corner (yellow circles). \textbf{a}, In the absence of loss, there are four distinct corner modes. \textbf{b}, Loss is implemented on two sublattices. This breaks the $C_{4v}$ symmetry of the system into $C_{2v}$ and leads to bimodes. \textbf{c}, By adding loss on three sublattices, the $C_{2v}$ symmetry is broken and the monomode emerges.}
 	\label{fig:2dexp} 	
\end{figure*}

\section{Comparision between theory and experiments}\label{sec:exp1}
\subsection{Experimental set-up}\label{sec:exp}
To experimentally validate the existence of monomodes, a photonic waveguide lattice was built. Although the theoretical model is quantum, it can be experimentally simulated using classical light. In this experimental setup, the propagation of the light is determined by the paraxial equation \cite{Longhi2009} 
\begin{equation}
\begin{split}
    i & \frac{\partial}{\partial z} \Psi(x, y, z)=\\
    &\left(-\frac{1}{2 k_0}\left(\frac{\partial^2}{\partial x^2}+\frac{\partial^2}{\partial y^2}\right)-\frac{k_0 \Delta n(x, y)}{n_0}\right) \Psi(x, y, z).
\end{split}
\label{eq_par}
\end{equation}
Here, $\Psi$ acts as the wavefunction for the electric field, $k_0 = n_0 \omega/ c$ , with $\omega$ the light frequency, and $n_0$ the refractive index, and $\Delta n$ is the change in refractive index. The paraxial equation has a similar structure to the Schrödinger equation. This system can be used to model the tight-binding Hamiltonian, where the hopping terms depend on the distance $d_i$ between the waveguides, $t_i \approx \mathrm{e}^{-d_i / \xi}$. $\xi$ depends on the parameters of the experiment, such as the wavelength of the light.
We achieve the desired loss in the system by introducing a certain concentration of microscopic scattering points along the waveguides through the dwelling process. The direct laser-writing technology provided us with the ability to freely tune both the dwelling time and the separation between the individual scattering points, allowing for the implementation of a wide range of artificial losses. Importantly, this process neither compromises the real part of the refractive index nor introduces directionality into the system. 
In addition, it offered us precise control over the amount and distribution of loss in the waveguide lattice, enabling us to design and fabricate the lattice with the desired characteristics.
To achieve this, a coherent light beam from a tunable laser (Cameleon Ultra II, Coherent) was launched into a glass sample using a 100× objective with a numerical aperture (NA) of 0.9. This configuration allows for individual excitation of each waveguide composing the structure. The output light of the glass sample was collected using a 20× objective, and the image profile of each individual waveguide forming the topological structure was captured using a CCD camera.

The topological photonic 3D waveguide lattice structures (2D spatial, 1D time) were fabricated using a pulsed femtosecond (fs) laser (BlueCut fs laser from Menlo Systems). The fs laser produced light pulses centered at a wavelength of $1030 \mathrm{~nm}$, with a duration of $350 \mathrm{fs}$, and a repetition rate of 1 $\mathrm{MHz}$. The waveguides were written in a Corning EAGLE2000 alumino-borosilicate glass sample with dimensions of $L=50, W=25$, and $h=1.1 \mathrm{~mm}$.

To inscribe the waveguide structures, pulses of $210 \mathrm{~nJ}$ were focused using a $50 \mathrm{X}$ objective of 0.55 NA. The waveguides were written at depths between 70 to $175 \mu \mathrm{m}$ under the surface, according to the designed structure, while the sample was translated at a constant speed of $30 \mathrm{~mm} / \mathrm{s}$ by a high-precision three-axis translation stage (A3200, Aerotech Inc.). The fabricated waveguides supported a Gaussian single mode at $780 \mathrm{~nm}$, with a mode field diameter $\left(1 / e^2\right)$ of approximately $6-8 \mu \mathrm{m}$. The mechanism of ultrafast laser pulses-material interaction gave a refractive index increase of about $2 \times 10^{-4}$. The propagation losses were estimated to be around $0.3 \mathrm{~dB} / \mathrm{cm}$, and the birefringence was in the order of $7 \times 10^{-5}$.

\subsection{1D SSH model}
To illustrate our findings, we now perform a set of eight experiments using waveguides for the 1D model. The first experiment (i) realizes an equally spaced tight-binding chain where the distance between the waveguides was set to $d_1 = d_2 = 10\ \mu m$. The next four experiments realize a dimerized chain with $d_1 = 12\ \mu m$ and $d_2= 10\ \mu m$ (or vice-versa) in the (ii) trivial limit, (iii) topological limit without loss, and (iv-v) topological limit with loss. This yields a ratio of $t_2/t_1\approx 2$ for the topological case. The last three experiments (vi-viii) realize topological defects. In this case, there is an edge mode at the end of the chain and one at the defect. Here, distances of $d_1=  11\ \mu m$ and $d_2=\  7\ \mu m$ were used, giving a ratio of $t_2/t_1\approx 3.2$. The loss for the fourth and fifth experiment was engineered by making scattering points $0.2\ mm$ apart in the waveguide, by waiting $0.5\ s$ with the laser. For the last three (topological defect) experiments, the loss was engineered by adding 50 cuts of $350\ \mu m$ along the waveguide. To ensure that the edge states would not hybridise before the measurement, a system size of 10 unit cells was chosen.

Fig.~\ref{fig:exp} shows the experimental results, along with the theoretical predictions. Yellow circles are included to indicate the point of injection of light. White, dashed, circles represent the results of the tight-binding simulations. The radius of the circles is proportional to the relative intensity. In Appendix \ref{app:numpred}, we provide a side-by-side comparison between numerical and experimental results. In the uniform 1D chain and in the trivial phase of the SSH model, the light disperses into the bulk (Figs. \ref{fig:exp}a and \ref{fig:exp}b, respectively). In the topological phase without loss, there are two edge modes (Fig.~\ref{fig:exp}c). However, when we add loss on the $A$ sublattice near the left edge (see red dot in Fig.~\ref{fig:exp}d), we only observe one edge mode on the right-hand side ($B$ sublattice). When the loss is placed on the $A$ sublattice but far away from the left edge (Fig.~\ref{fig:exp}e), the left edge mode does not feel it immediately, and does not decay within the time scale of the experiment. This is in agreement with the theoretical predictions, represented by the white dashed circles (see Appendix \ref{app:numpred} for a stand-alone figure of the theoretical predictions). The selective elimination of one of the sublattices due to a single loss also occurs when there is an inhomogeneous hopping in the lattice. To show that, we create a chain with half of the dimerization corresponding to a topological phase, while the other half corresponds to a trivial phase. The site between the two chains is a kink, a kind of topological defect that can host fractionalized excitations in electron systems \cite{Teo2010}.  When the system hosts a topological defect in the center of the chain, there is an edge mode on the left edge of the chain, as well as a mode pinned on the defect (Fig.~\ref{fig:exp}f). By putting loss on the $B$ sublattice near the defect, the mode pinned on the defect can be destroyed, while leaving the left edge mode intact (Fig.~\ref{fig:exp}g). Similarly, the loss can be placed on the $A$ sublattice near the left edge, thus removing the edge mode, but leaving the monomode on the defect (Fig.~\ref{fig:exp}h). The existence of the topological mode at a defect is also understood in the framework of Jackiw-Rebbi theory \cite{Jackiw1976} using that the mass profile in the low-energy Dirac Hamiltonian changes sign across the defect, as discussed in Sec.~\ref{sec:JR}.

\subsection{2D SSH model}
For the 2D model, three experiments were realized. The distance between the waveguides was set to $d_1= 11\ \mu m$ and $d_2= 7\ \mu m$, which yields a ratio of $t_2/t_1\approx 3.2$. The loss was engineered by adding 100 cuts of $70\ \mu m$ along the waveguide. A system of 5 by 5 unit cells was used.

Fig.~\ref{fig:2dexp} shows the results of the experiment for the 2D model. Yellow circles are included to signal the point of injection of light. In the lossless case, the system hosts four corner modes (Fig.~\ref{fig:2dexp}a). By adding loss to two sublattices, the $C_4$ symmetry is reduced to $C_{2v}$ and two of the corresponding corner modes are destroyed, leaving bimodes (Fig.~\ref{fig:2dexp}b). Adding loss to one more sublattice breaks the $C_{2v}$ symmetry further and reveals a monomode in the 2D SSH model (Fig.~\ref{fig:2dexp}c). Notice that it is not possible to create a monomode by only applying loss on a single lattice site for this model. The 2D SSH model has four distinct sublattices, so three lossy sites at different sublattices are required to obtain a monomode. This is in full agreement with the theoretical predictions (see Appendix \ref{app:numpred}).

\begin{figure*}
	\centering
	\includegraphics{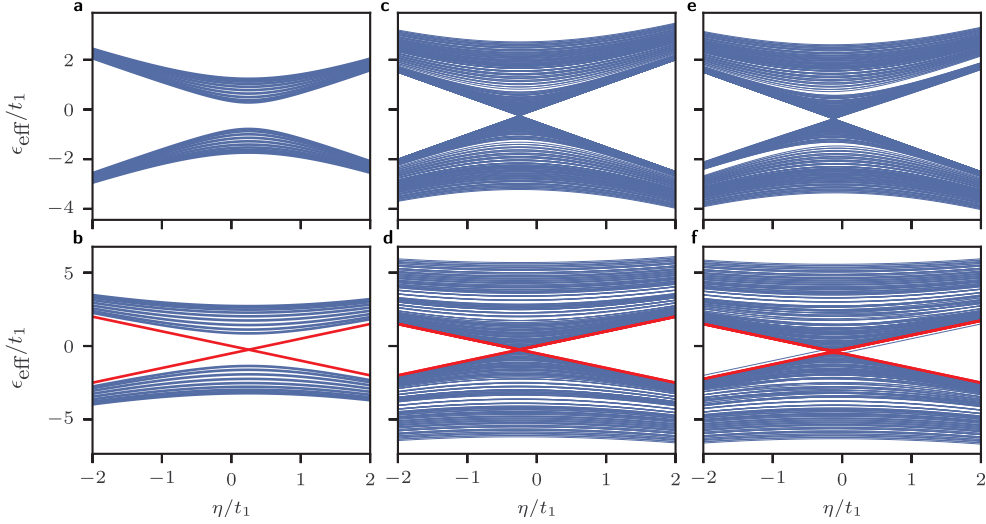}
	\caption{Spectrum of the effective Hamiltonian for the 1D and 2D models. \textbf{a}, Trivial phase for a 1D NH SSH model with staggered loss; \textbf{b}, topological phase. The linearly dispersing topological modes in the gap are shown in red. \textbf{c}, Spectrum of the 2D NH SSH model with the loss configuration sketched in Fig.~\ref{fig:2dexp}b in the trivial phase and \textbf{d} in the topological phase. \textbf{e} Spectrum of the 2D SSH model with the loss configuration sketched in Fig.~\ref{fig:2dexp}c in the trivial phase and \textbf{f}, in the topological phase. All spectra in the trivial phase are calculated for ($t_2/t_1=0.5$), and in the topological phase, for ($t_2/t_1=2$). Modes with an absolute value of the wavefunction larger than 0.8 at the ends (for 1D) and corners (for 2D) of the lattice are plotted in red; other modes are plotted in blue. We used 10-unit cells in each direction for this simulation.}
	\label{fig:topology}
\end{figure*}
\section{Topological analysis}\label{sec:topo}
To characterize the topology of this model, we first consider how the addition of loss in the SSH chain changes its symmetries. Even though sublattice symmetry is broken by the presence of on-site loss, (non-Hermitian) chiral symmetry $\Gamma:\Gamma H \Gamma = - H^\dagger$ is still present for alternating loss. In general, models that have this symmetry are of the form
\begin{equation}
	H(k) = \mqty(iP(k) & Q(k) \\ Q^\dagger(k) & i R(k)),
\end{equation}
where $P(k)$ and $R(k)$ are Hermitian matrices and $Q$ is non-Hermitian in general. Notice that the constraint that a chiral Hermitian Hamiltonian should be block off-diagonal is relaxed when considering non-Hermitian Hamiltonians.

For the NH SSH model with on-site loss on one sublattice, we have $\Gamma=\sigma_z$. As a result of the chiral symmetry, it is possible to characterize the phases of the non-Hermitian SSH model in terms of a topological invariant. The calculation of this invariant is based on defining a Hermitian ancestor Hamiltonian following Refs. \cite{Miri2019,Brzezicki2019, Hyart2022},
 \begin{equation}
 	H^\text{eff}(\eta) = \Gamma (\eta\mathbb{I}-iH),
 	\label{eq:hameff}
 \end{equation}
where $\mathbb{I}$ is the 2D identity matrix. Notice that $\eta$ acts as an extra (continuum) momentum for the Hamiltonian, which makes that this effective Hermitian Hamiltonian has always a higher dimension than the corresponding non-Hermitian one.

The Bloch Hamiltonian for the NH SSH model with loss on the $A$ sublattice is given by
 \begin{equation}
     H_{NH-SSH}(k) = \begin{pmatrix}
         -i\gamma & t_1 + t_2 e^{ik} \\
         t_1 + t_2 e^{-ik} & 0
     \end{pmatrix}.
     \label{eq:H_SSH}
 \end{equation}
From Eq.~\eqref{eq:hameff}, we then obtain the following effective Hermitian Hamiltonian: 
\begin{equation}
    H^\text{eff}(k,\eta) =\begin{pmatrix}
         \eta-\gamma & -i\left(t_1 + t_2 e^{ik}\right)\\
         i\left(t_1 + t_2 e^{-ik}\right)&-\eta
     \end{pmatrix}. \label{eq:H_eff_1D_SSH}
\end{equation}
\subsection{Chern number}
The effective Hamiltonian $H^{\text{eff}}$ for the non-Hermitian SSH model given by Eq.~\eqref{eq:H_SSH} corresponds to a Chern insulator, such that we can associate a Chern number to it. This model fully captures the trivial and topological phases of the non-Hermitian SSH model with loss in only one sublattice, which is corroborated by the spectra shown in Figs. \ref{fig:topology}a and \ref{fig:topology}b, respectively. For the trivial phase, Fig.~\ref{fig:topology}a, the model exhibits a completely gapped energy spectrum, whereas for the topological phase, Fig.~\ref{fig:topology}b, two linearly dispersing midgap states arise. These are chiral edge states, present in the topological phase of a Chern Insulator. Using the effective Hermitian Hamiltonian, we can also calculate the Berry curvature and the Chern number, which is quantized as long as the Hamiltonian is periodic in both $k$ and $\eta$. However, here the Hamiltonian is periodic in the wave number k, but anti-periodic in $\eta$: $H^\text{eff}(k,\eta\to -\infty) = - H^\text{eff}(k,\eta\to \infty)$. To circumvent this problem, we perform the following, unitary transformation to obtain the compactified effective Hamiltonian 
 \begin{equation}
     H^\text{eff}_\text{cp}(k,\eta) = \mathcal{R}_\eta H^\text{eff}(k,\eta) \mathcal{R}^\dagger_\eta,
 \end{equation}
 with
 \begin{equation}
     \mathcal{R}_\eta = \exp \left[ i \frac{\pi}{4} (1 + \tanh{\eta}) \begin{pmatrix}
         0 & 1 \\ 1 & 0
     \end{pmatrix} \right].
 \end{equation}
Since the transformation that we performed is unitary, the spectra of $H^\text{eff}_\text{cp}(k,\eta)$ and $H^\text{eff}(k,\eta)$ are equivalent. 
 \begin{widetext}
\noindent The Chern number for $H^\text{eff}_\text{cp}(k,\eta)$ is now obtained through
 \begin{equation}
     C = \frac{1}{2\pi} \int_{-\infty}^\infty \dd \eta \int_0^{2\pi} \dd k\ \Omega_{k,\eta},
 \end{equation}
 with the Berry curvature $\Omega_{k,\eta}$ given by
 \begin{equation}
     \Omega_{k,\eta} = 2\sum_{n\leq n_F,m > n_F} \Im \frac{\bra{\psi^{n}_{k,\eta}}\partial_k H^\text{eff}_\text{cp} \ket{\psi^m_{k,\eta}}\bra{\psi^{m}_{k,\eta}}\partial_\eta H^\text{eff}_\text{cp} \ket{\psi^n_{k,\eta}}}{\left( E_{k,\eta}^{(n)} -  E_{k,\eta}^{(m)} \right)^2}.
     \label{eq:method:omega}
 \end{equation}
 Here, $E_{k,\eta}^{(i)}$ and $\ket{\psi^i_{k,\eta}}$ are the eigenvalues and normalized eigenstates of the $i$-th band of $H^\text{eff}_\text{cp}$, and $n_F$ is the number of occupied bands. It turns out that calculating the Chern number is easier for the non-Hermitian SSH model with staggered loss. In this case, the eigenvalues of $H^\text{eff}_\text{cp}$ are given by
 \begin{equation}
     E^\pm_{k,\eta} = \gamma \pm \sqrt{t_1^2 + t_2^2 + \eta^2 +2 t_1 t_2 \cos k },
 \end{equation}

while the eigenvectors are given by 
 \begin{equation}
         \ket{\psi^\pm_{k,\eta}} =
     \mathcal{N}_{\pm,k}\begin{pmatrix}
           \pm  i\frac{\sqrt{t_1^2 + t_2^2 + \eta^2 +2 t_1 t_2 \cos k } \mp (t_1+t_2 \cos k) \cos \left( \frac{\pi}{2} \tanh \eta \right) \mp \eta \sin \left( \frac{\pi}{2} \tanh \eta \right) }{-\eta \cos \left( \frac{\pi}{2} \tanh \eta \right) + i t_2 \sin k + (t_1+t_2 \cos k) \sin \left( \frac{\pi}{2} \tanh \eta \right) } \\ 1
     \end{pmatrix},
 \end{equation}
 \end{widetext}
 where $\mathcal{N}_{\pm,k}$ is a normalization factor. 
 
 We observe that the Berry curvature in Eq.~\eqref{eq:method:omega} is independent of $\gamma$. Moreover, we obtain
 \begin{equation}
     C = \left\{
 \begin{array}{ll}
       0 & |t_1| > |t_2| \\
       1 & |t_1| < |t_2| \\
 \end{array} 
 \right..
 \end{equation}
Notice that this bulk invariant was computed for the system with alternating loss because it is a translation invariant version of the model with a single lossy site. Both exhibit the same topological properties since they belong to the same symmetry class. The nontrivial Chern number marks the topological character of the chiral edge states present in $H^\text{eff}$. It labels the topology of both edge modes, although only the monomode survives for long time.

\subsection{Jackiw-Rebbi}\label{sec:JR}
The methodology of the effective Hamiltonian can also be used to investigate both a topological defect in the 1D SSH model and corner modes in the 2D SSH model. We performed calculations using the Jackiw-Rebbi method \cite{Benalcazar2017, Jackiw1976}. This method is based on the derivation of a low-energy effective Dirac Hamiltonian for the system. In this description, the topology of the system is determined by the spatial structure of the mass term. A change in the mass term from the topological to the trivial phase yields localized zero-energy states.

Writing the effective Hamiltonian [Eq.~\eqref{eq:H_eff_1D_SSH}] in terms of Pauli matrices $\sigma_i,$
 \begin{equation}
     \begin{split}
        H^\text{eff}(k,\eta)&=\left(\eta-\frac{\gamma}{2}\right)\sigma_z-\frac{\gamma}{2}\mathbb{I}+\left[t_1+t_2\cos(k)\right]\sigma_y\\
     &\phantom{=}+t_2 \sin(k)\sigma_x,         
     \end{split}
     \label{eq:go}
 \end{equation}
one observes that the gap closing occurs for $t_1=t_2$, $k=\pm \pi$ and $\eta=\gamma/2$. Close to the phase transition and in the vicinity of the gap closing point, we can perfom a low-energy description of this model
 \begin{equation}\begin{split}
     H^\text{eff}(p,\eta)&=\left(\eta-\frac{\gamma}{2}\right)\sigma_z+t_2 M\sigma_y \\
     &\phantom{=} - t_2 p\sigma_x+\mathcal{O}(p^2),
 \end{split}\end{equation}
 where $p=k-\pi$, the mass term $M=\left(t_1-t_2\right)/t_2$, and we changed the energy reference to neglect the term proportional to the identity. The quadratic term can be added to fix the boundary conditions. The gap closing then occurs for $p=0$ and $M=0$, and the sign of the mass term determines whether the system is trivial ($M>0$) or topological ($M<0$). Associated with this low-energy model, there is a continuum model
 \begin{equation}
     H^\text{eff}(x,\eta)=\left(\eta-\frac{\gamma}{2}\right)\sigma_z+t_2 M\sigma_y+i t_2 \partial_x\sigma_x,
 \end{equation}
 since the momentum is the derivative $p=-i\partial_x$ along the coordinate $x$ of the lattice.

 Using the continuum Hamiltonian, a change in the hopping parameters along the lattice is translated into a spatially dependent mass term $M(x)$. A topological defect is a domain wall between a topological ($M<0$) and a trivial  ($M>0$) lattice, so we can model it by a profile of mass that changes sign across the boundary and vanishes at the defect. Therefore, for at least one value of $\eta$, there will be a localized zero-energy solution of the Hamiltonian
 \begin{equation}\begin{split}
     H^\text{eff}\left(x,-\frac{\gamma}{2}\right)\Psi_{DW}(x)&=\left[t_2 M(x)\sigma_y+i t_2 \partial_x\sigma_x\right]\Psi_{DW}(x) \notag \\ 
     =0&=\left[M(x)\mathbb{I}+\partial_x\sigma_z\right]\Psi_{DW}(x).
 \end{split}\end{equation}
Notice that we multiplied the matrix applied on  $\Psi_{DW}(x)$ by $\sigma_y$ and divided by $t_2$ to obtain the last equality.

$\Psi_{DW}(x)$ may be expressed in terms of the eigenstates of $\sigma_z$, $\chi^{+}=\begin{pmatrix} 1&0\end{pmatrix}$ and $\chi^{-}=\begin{pmatrix} 0&1\end{pmatrix}$,
 \begin{equation}
     \Psi_{DW}(x)=\sum\limits_{\sigma=\pm 1}c_\sigma\psi^{\sigma}_{DW}(x)\chi^\sigma,
     \label{eq_Psi_DW}
 \end{equation}
 where
 \begin{equation}
    \sum\limits_{\sigma}\left|c_\sigma\right|^2=1, \quad \int\limits_{-\infty}^{\infty} dx \left|\psi_{DW}^\sigma(x)\right|^2=1.
 \end{equation}
 We can obtain a differential equation for $\psi_{DW}^\sigma(x)$ in terms of the eigenvalues $\sigma$ of $\sigma_z$,
 \begin{equation}
    \left[M(x)+\sigma\partial_x\right]\psi_{DW}^\sigma(x)=0,
\end{equation}
such that
\begin{equation}
    \psi_{DW}^\sigma(x)=\mathcal{N} \exp{-\sigma \int\limits_{-\infty}^{x} M(x') dx'},
 \end{equation}
 where $\mathcal{N}$ is a normalization factor. For a system with the topological phase on the left and the trivial phase on the right of the domain wall, we obtain localized solutions just for $\sigma=1$; therefore, $c_-=0$ and $c_+=1$. Notice that at the domain wall, the Hamiltonian as a function of $\eta$ is given by
 \begin{equation}
     H^\text{DW}(\eta)=\langle\Psi_{DW}|H^\text{eff}|\Psi_{DW}\rangle=\eta-\frac{\gamma}{2},
 \end{equation}
 which exhibits a linear dispersion. Moreover, there is a localized solution only at the boundary between the trivial ($M>0$) and topological ($M<0$) phases, which characterizes this defect as a topological defect.

The same method can be used to show the topological characteristic of the corner modes in the 2D SSH model. Even though this model does not present a gapped bulk spectrum, zero energy modes localized in the corner only appear at the HOT phase. We start by extending the method from Refs.~\cite{Brzezicki2019, Hyart2022} to non-Hermitian 2D systems. For the 2D SSH model, it takes the form (Appendix \ref{app:nonhem1}) 
\begin{equation}
	H^\text{eff}(\mathbf{k}, \eta) = \Gamma\qty(\eta\mathbb{I} \otimes \mathbb{I} - i H_\text{NH-2D-SSH}(\mathbf{k})),
\end{equation}
where now the chiral symmetry is $\Gamma \equiv \sigma_z \otimes \mathbb{I}$ since we are dealing with a four orbital model. Although there is no longer a Chern number associated with this model, it is still useful to analyse the topological properties of boundary states (Appendix \ref{app:nonhem2}). We obtain the low-energy Hamiltonian 
\begin{equation}\begin{split}
	H^\text{eff}(\mathbf{p}, \eta) &= H_d(\eta) + t_2M^x\sigma_y\otimes\mathbb{I} - t_2p_x\sigma_x\otimes\sigma_z\\
    &\phantom{=}+t_2M^y\sigma_y\otimes\sigma_x + t_2p_y\sigma_y\otimes\sigma_y,
\end{split}\end{equation}
with $p_x = k_x-\pi$ and $p_y=k_y-\pi$. Again, the topological phase is characterized by the mass terms $M^{x/y}$. To model the different boundaries, we set $t_1$ to be different across $x$ and $y$ and define $M^{x/y}=(t_1^{x/y}-t_2)/t_2$. The HOT phase is marked by negative $M^x$ and $M^y$, while we model the exterior of the lattice by a trivial insulator with positive $M^{x/y}$. By performing a calculation analogous to the one done for the topological defect, we show in the SM that one obtains localized states at the corners only when there is this change of sign for the mass term, indicating the topological aspect of these zero-energy boundary modes. These modes also appear as linearly dispersing edge modes for the effective Hamiltonian of Eq.~\eqref{eq:hameff} as shown in Figs. \ref{fig:topology}c-f. For the loss configuration of Fig.~\ref{fig:2dexp}b, the spectrum (Fig. \ref{fig:topology}c) of the trivial phase does not display modes localized in the corners, while the one for the topological phase does (Fig. \ref{fig:topology}d). The same holds for the loss configuration of Fig.~\ref{fig:2dexp}c, see Figs.~\ref{fig:topology}e-f.

\section{Conclusions \& Outlook}\label{sec:conc}
We have theoretically and experimentally shown the morph of non-Hermitian symmetries through the engineering of loss. By introducing losses on a selected sublattice and sufficiently close to the corresponding edge, one of the topological edge modes decays over time. We have realized these monomodes experimentally in a photonic lattice. Moreover, we confirmed the robustness of the monomodes against perturbations. It is remarkable that a generalized topological invariant protecting the corner mode remains valid in this non-Hermitian setup. 

The monomode concept is related to an intriguing property experimentally demonstrated here, in which one can add an NH term that breaks the original Hermitian symmetry of the model but preserves a generalized NH symmetry. In this work, this NH perturbation is staggered loss, which explicitly breaks the Hermitian sublattice symmetry of the SSH model while preserving NH chiral symmetry. We have shown that the topology of this model can be understood in terms of an effective Hermitian Hamiltonian in higher dimension. Although this is constructed explicitly here for chiral symmetry, we speculate that this method can be extended to other symmetries. 

The implications of our results are multifold. On the one hand, we have identified an extremely simple model, capable of revealing monomodes. Since those cannot recombine with their corresponding partner, those monomodes would be ideal candidates for transmitting information based on topological states. Here, we have the additional advantage that this mode will not hybridize, and hence exhibits further robustness with respect to its Hermitian counterpart. In this context, it is important to note that the phenomena explored in this work generalizes, mutatis mutandis, to quantum master equations \cite{Hegde2023, Yang2023}. A natural extension of this work is to explore $\mathbb{Z}_2$ topological classes and investigate whether the engineered loss in another inner degrees of freedom (like spin or particle/hole) can lead to the morphing of symmetries. 

\subsection*{Acknowledgement} 
The authors thank J.C. Budich, J. Carlström, T. Nag, A. Black-Schaffer, and J.A. Klärs for useful discussions. WC, RA, EJB, and MB thank the Knut and Alice Wallenberg Foundation for financial support (grant numbers 31002636, 2018.0071, and 31001206). WC, EJB, and MB thank the Swedish Research Council for financial support (grant numbers 2017-04855 and 31001205). LE and CMS acknowledge the research program “Materials for the Quantum Age” (QuMat) for financial support. This program (registration number 024.005.006) is part of the Gravitation program financed by the Dutch Ministry of Education, Culture and Science (OCW). 

\subsection*{Author contribution}
ES did almost all theoretical calculations under the supervision of LE, RA, and CMS. The Chern numbers were calculated by LE, and the topology in the framework of the Jackiw-Rebbi theory was calculated by RA under the supervision of EJB. The experiments were performed by WC under the supervision of MB. CMS coordinated the entire project. All authors discussed the results and contributed to the elaboration of the manuscript. 

\appendix

\section{Gain and Loss Systems}
\label{app:gainandloss}
Here, we investigate different loss configurations in the SSH model. In this case, the non-Hermiticty will be caused by gain and/or loss on the sites. We will start by looking at the dimer with both gain and loss, followed by the SSH model with gain and loss. Henceforth, we will study a system with only loss.

\subsection{Broken $\mathcal{PT}$-symmetry}
The systems discussed here can exhibit $\mathcal{PT}$-symmetry, depending on the system size and choice of parameters. $\mathcal{PT}$-symmetry is the invariance of a system to combined parity ($x \rightarrow -x$) and time inversion ($t\rightarrow -t$) transformations. To illustrate how this works, we will discuss a very small and simple system showing both a $\mathcal{PT}$-conserved phase and a spontaneously broken phase. We note that the same reasoning works for larger lattices. The system consists of two sites, with gain $i\gamma$, loss $-i\gamma$, and hopping $t_1$ between the sites. The Hamiltonian reads
\begin{equation}
	H = \mathbf{c}^\dagger \mqty(-i \gamma & -t_1 \\ -t_1 & i \gamma)\mathbf{c},
	\label{eq:hammatsystem1open}
\end{equation}
where $\mathbf{c}$ ($\mathbf{c}^\dagger$) are vectors of annihilation (creation) operators. The Hamiltonian of Eq.~\eqref{eq:hammatsystem1open} is clearly non-Hermitian and $\mathcal{PT}$-symmetric. Its eigenvectors and eigenvalues are given by
\begin{equation}
	\phi_{\pm} = \mqty(-\frac{-i\gamma \pm \sqrt{t_1^2-\gamma^2}}{t_1} \\ 1) \qqtext{and} \epsilon_{\pm} = \pm\sqrt{t_1^2-\gamma^2}.
\end{equation}
There is an exceptional point at $\gamma = t_1$ because at this point both eigenvalues are 0 and both eigenvectors are $\phi = \mqty(i & 1)^T$. For $\gamma < t_1$, the spectrum has fully real eigenvalues, but for $\gamma > t_1$ it has fully imaginary eigenvalues, in conjugated pairs. This suggests the $\mathcal{PT}$-symmetry is broken for $\gamma>t_1$, which can be proven considering the eigenvectors of the system. For $\gamma < t_1$, we have
\begin{equation}
	\mathcal{PT}\phi_\pm = \mathcal{P}\mqty(-\frac{i\gamma \pm \sqrt{t_1^2-\gamma^2}}{t_1} \\ 1)=\mqty(1\\ -\frac{i\gamma \pm \sqrt{t_1^2-\gamma^2}}{t_1}).
\end{equation}
Now, we multiply the resulting vector by a constant to obtain
\begin{equation}
	\mathcal{PT}\phi_\pm = \mqty(-\frac{t_1}{i\gamma \pm \sqrt{t_1^2 - \gamma^2}}\\1) = \mqty(-\frac{-i\gamma \pm \sqrt{t_1^2-\gamma^2}}{t_1} \\ 1) = \phi_\pm,
\end{equation}
so the eigenvectors are indeed $\mathcal{PT}$-symmetric. When $\gamma > t_1$, the eigenvectors are equal to
\begin{equation}
	\phi_{\pm} = \mqty(-\frac{-i\gamma \pm i \sqrt{\gamma^2-t_1^2}}{t_1} \\ 1)
\end{equation}
and when we now apply the $\mathcal{PT}$ operators to it, we obtain
\begin{equation}
\begin{split}
	&\mathcal{PT}\mqty(-\frac{-i\gamma \pm i \sqrt{\gamma^2-t_1^2}}{t_1} \\ 1) \\
 &= \mathcal{P}\mqty(-\frac{i\gamma \mp i \sqrt{\gamma^2-t_1^2}}{t_1} \\ 1) \\
 &= \mqty(1\\-\frac{i\gamma \mp i \sqrt{\gamma^2-t_1^2}}{t_1}) \\
 &\rightarrow \mathcal{PT}\phi_\pm = \mqty(-\frac{t_1}{i\gamma \mp i \sqrt{\gamma^2-t_1^2}}\\1) \neq \phi_\pm.
\end{split}
\end{equation}
We indeed find that the eigenvectors are no longer $\mathcal{PT}$-symmetric if $\gamma > t_1$ and thus, the eigenvalues become complex. However, when the system size increases, this $\mathcal{PT}$-symmetric region decreases in size, such that the $\mathcal{PT}$-symmetric phase is absent in the thermodynamic limit.

\subsection{Alternating Gain and Loss}
\begin{figure*}[t]
	\center
    \includegraphics[]{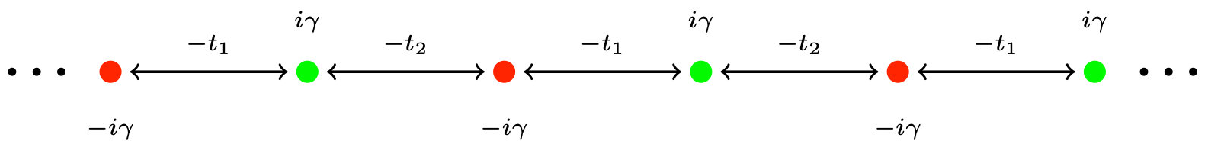}
	\caption{SSH chain with alternating gain and loss.}
	\label{fig:gainandlosschain}
\end{figure*}
We now consider a reciprocal SSH chain with alternating gain and loss on the sites \cite{Roccati2022, Miri2019}, a system that has been extensively discussed in the literature, see Fig.~\ref{fig:gainandlosschain}. The reason why these imaginary contributions on the site lead to gain or loss can be understood straightforwardly by looking at the time propagator,
\begin{equation}
	U(t) = e^{-\frac{i}{\hbar}\hat{H}t},
\end{equation}
where $t$ is time. By acting with this operator on an eigenstate of $H,\ $$\ket{\psi_n}$, the state will evolve in time according to its eigenvalue
\begin{equation}
	U(t)\ket{\psi} = e^{-\frac{i}{\hbar} \epsilon_n t}\ket{\psi_n}.
\end{equation}
When there are losses at the sites, the eigenvalues acquire a negative imaginary contribution. This will cause an exponential decay of the state, meaning loss. In the same way, positive imaginary contributions to the eigenvalues signify gain. The matrix Hamiltonian of this system is given by
\begin{equation}
	 H =  \mathbf{c}^\dagger\begin{pmatrix}
 	-\gamma i& -t_1 & 0 & \cdots & b\\ 
 	-t_1 & \gamma i & -t_2 &  & \\ 
 	0 & -t_2 & -\gamma i &  & \vdots \\ 
 	\vdots&  &  & \ddots  & \\ 
 	 b&  & \cdots &  & \gamma i
	\end{pmatrix}\mathbf{c},
	\label{eq:hammatgainloss}
\end{equation}
where $b$ is $-t_2$ or zero for periodic (PBC) or open boundary conditions (OBC), respectively. After diagonalizing the matrix we get,
\begin{equation}
	\epsilon(k) = \pm\sqrt{-\gamma^2 + t_1^2 + t_2^2 + 2 t_1 t_2 \cos(ka)}.
	\label{eq:spectrumgainloss}
\end{equation}
We can now study this system in the topological and trivial regime.

\begin{figure*}[t]
    \centering
    \includegraphics{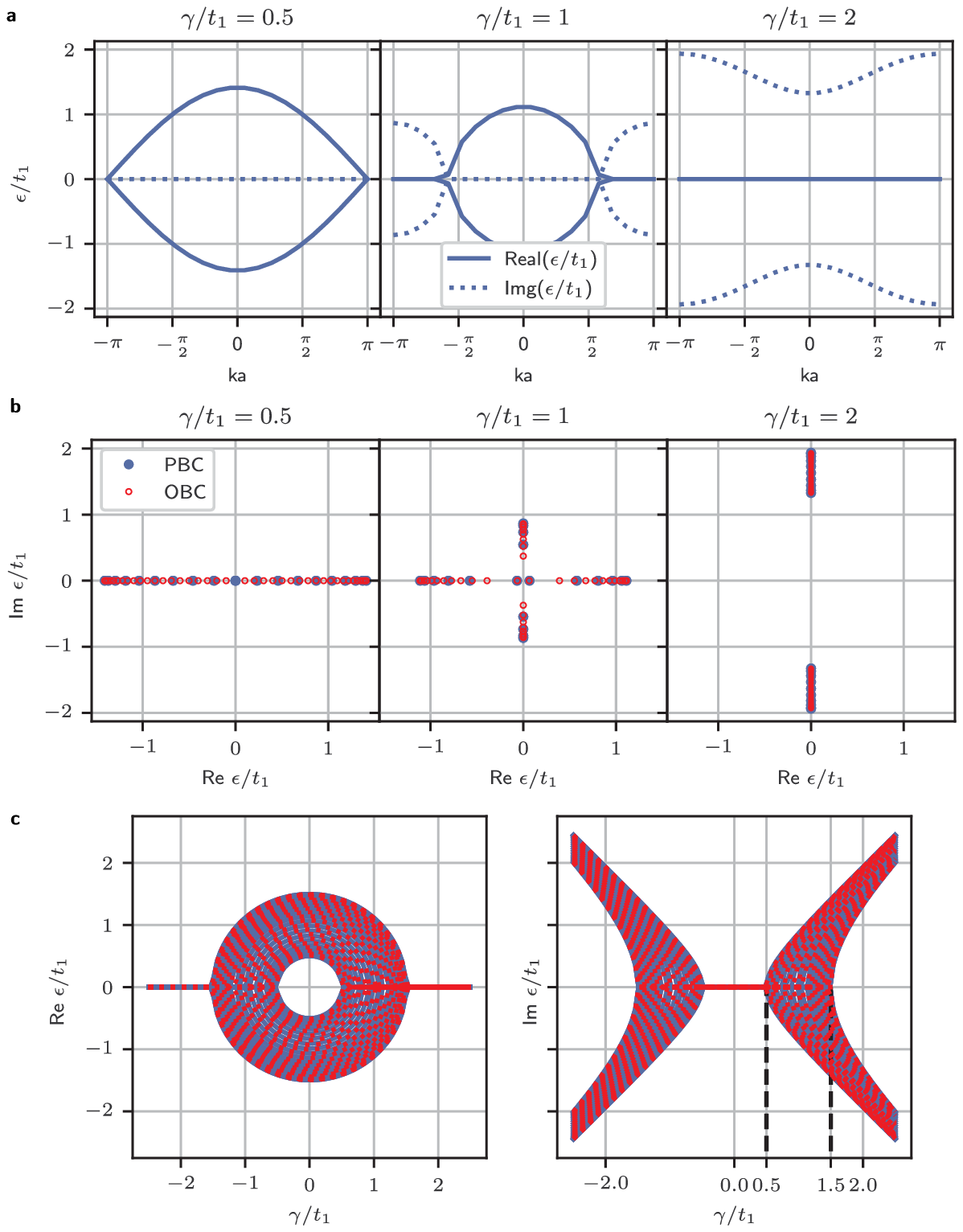}
    \caption{Spectrum of the SSH model with alternating gain and loss for PBC and OBC in the trivial regime $t_2 = t_1/2$. A system size of 20 unit cells was used. \textbf{a}, As a function of $ka$ for different values of $\gamma$. \textbf{b}, In the complex plane for different values of $\gamma$. The blue full circles results correspond to PBC, and the red empty circles to OBC. \textbf{c}, As a function of $\gamma/t_1$.}
    \label{fig:gl-trivial}
\end{figure*}

\subsubsection{Trivial Phase}
In the trivial regime of the SSH model, $\abs{t_1/t_2} \leq 1$, the system has a fully real spectrum for $|\gamma| \leq |t_1-t_2|$ because the eigenvectors of the Hamiltonian remain $\mathcal{PT}$-symmetric. This can be seen in Figs.~\ref{fig:gl-trivial}a-b, where the imaginary part is zero for $|\gamma| < |t_1 - t_2|$ ($|t_1 - t_2| = 0.5$ as $t_1 = 1$ and $t_2 = 0.5$). In this regime, the system also has a (real) band gap. The system opens up a line gap for $|\gamma| > |t_1 + t_2|$. A line gap is present when one can draw a line between two segments of the spectrum. This is different to a point gap, where the spectrum is encircling a point. Different representations of the spectrum can be seen in Figs.~\ref{fig:gl-trivial}a-c. In the case of OBC, the spectra of the system will not change much besides finite-size effects, being equal to the PBC one in the thermodynamic limit.

\begin{figure*}[t]
    \centering
    \includegraphics{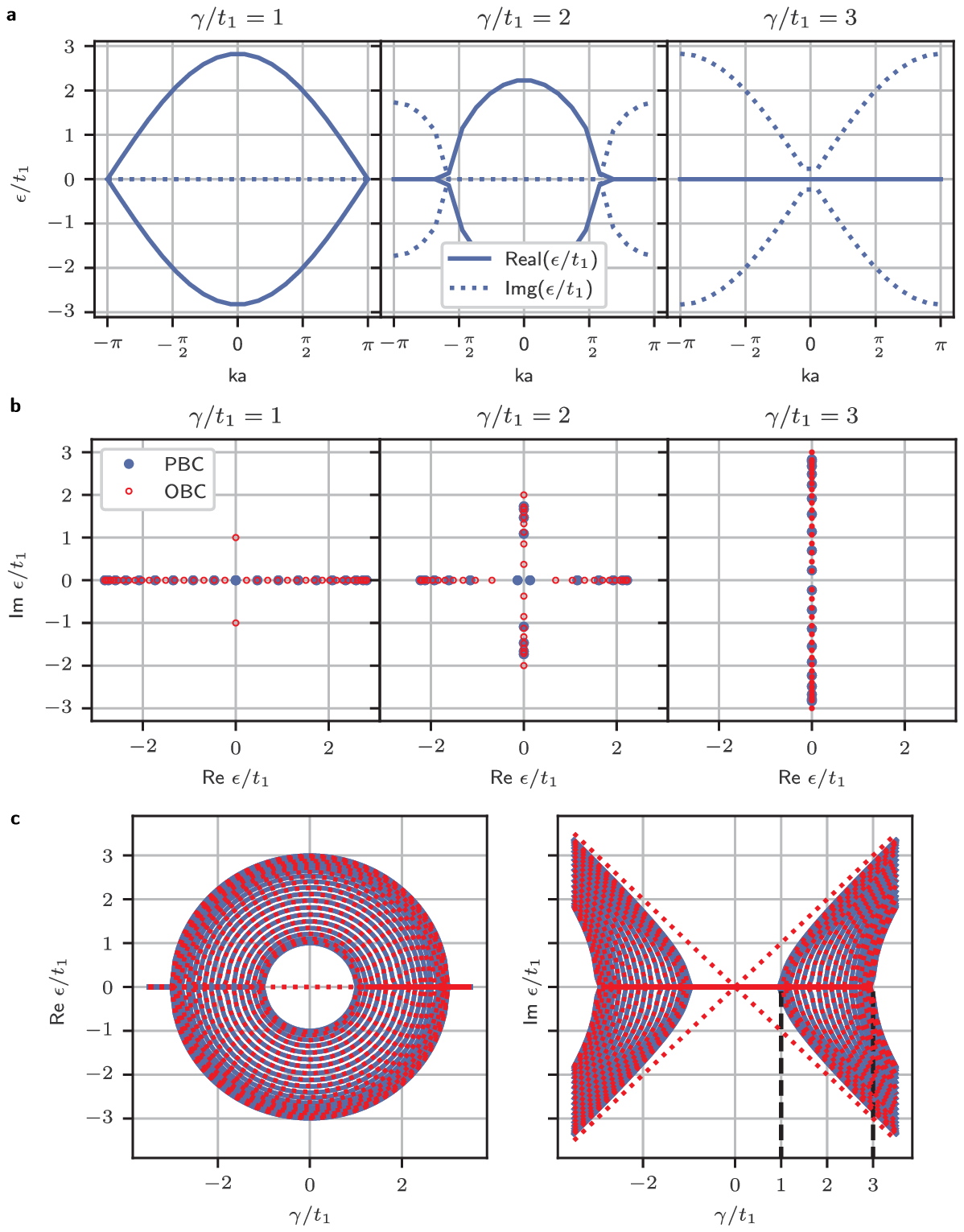}
    \caption{Spectrum of the SSH model with alternating gain and loss for PBC and OBC in the topological regime $t_2 = 2 t_1$. A system size of 20 unit cells was used. \textbf{a}, As a function of $ka$ for different values of $\gamma$. \textbf{b}, In the complex plane for different values of $\gamma$. The blue line results correspond to PBC, and the red line to OBC. \textbf{c}, As a function of $\gamma/t_1$.}
    \label{fig:gl-topo}
\end{figure*}

\subsubsection{Topological Phase}
The topological regime of the SSH model exists when $\abs{t_2/t_1} > 1$. In this section, we will specifically consider the case when $t_2 = 2 t_1$. For PBC, the system behaves in the same way as in the trivial regime. $\mathcal{PT}$-symmetry is conserved for $|\gamma| \leq |t_1-t_2|$ and the system opens up a line gap for $|\gamma| > |t_1+t_2|$. Different representations of the spectrum can be seen in Figs.~\ref{fig:gl-topo}a-c.
The topological regime of the SSH model is more interesting in the OBC case, in which there are topologically-protected midgap states. In this non-Hermitian SSH model, these edge modes will no longer have zero energy. In Fig.~\ref{fig:gl-topo}b, the spectra in the imaginary plane is given for  specific parameters. The zero energy modes have each acquired an imaginary energy. One mode acquired $+i\gamma$, while the other got $-i\gamma$. These zero modes are also clearly visible in Fig.~\ref{fig:gl-topo}c, where they are the cause of the $\mathcal{PT}$-symmetry breaking for all values $\abs{\gamma}>0$.

\subsection{Alternating Loss}
\begin{figure*}[t]
	\center
    \includegraphics[]{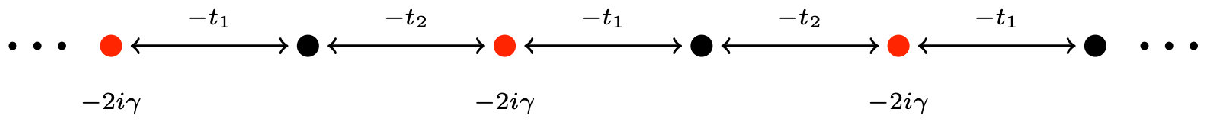}
	\caption{SSH chain with alternating loss.}
	\label{fig:losschain}
\end{figure*}

\begin{figure*}[t]
    \centering
    \includegraphics{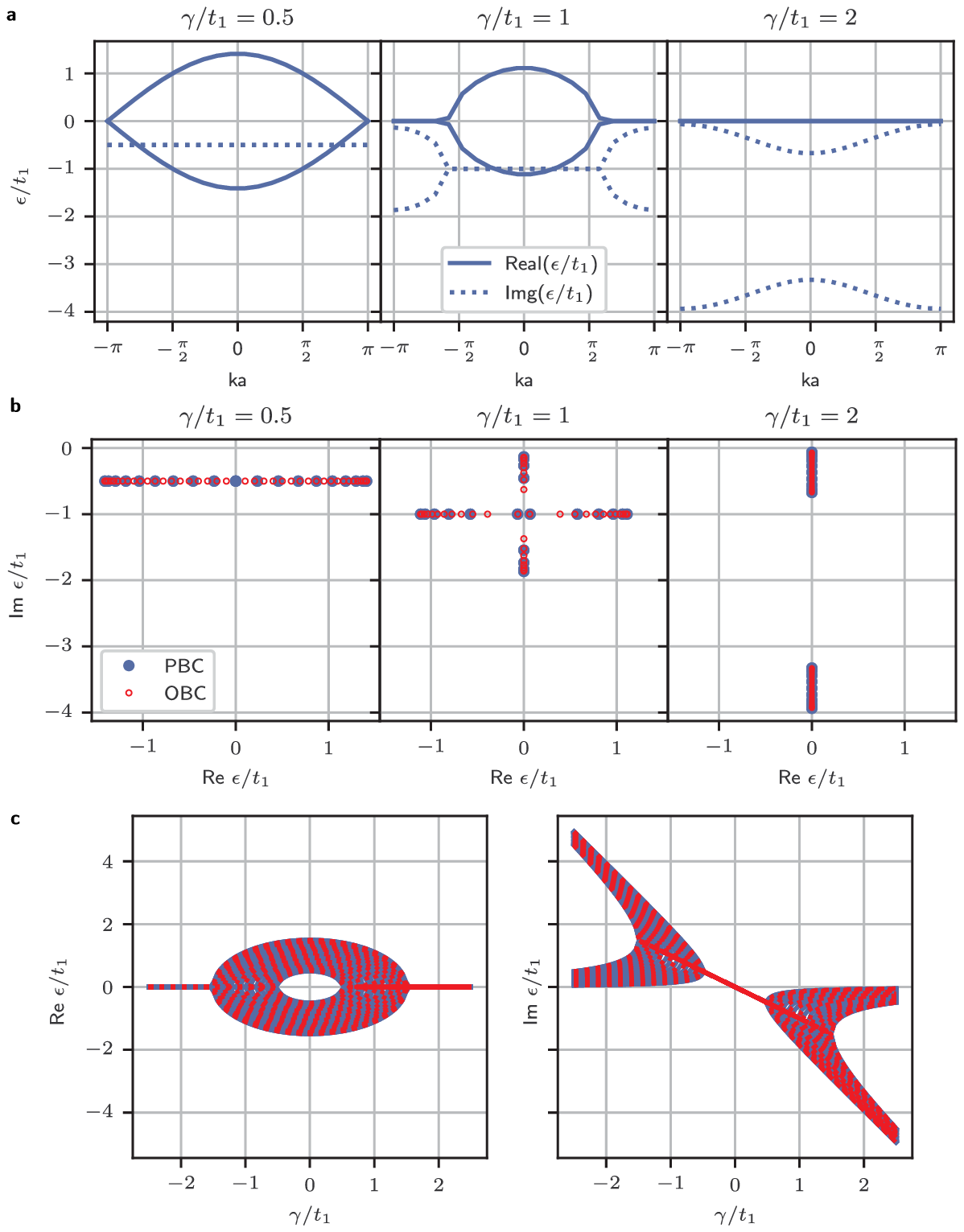}
    \caption{Spectrum of the SSH model with alternating loss for PBC and OBC in the trivial regime $t_2 = t_1/2$. A system size of 20 unit cells was used. \textbf{a}, As a function of $ka$ for different values of $\gamma$. \textbf{b}, In the complex plane for different values of $\gamma$. The blue full circles results correspond to PBC, and the red empty circles to OBC. \textbf{c}, As a function of $\gamma/t_1$.}
    \label{fig:l-trivial}
\end{figure*}
\begin{figure*}[t]
    \centering
    \includegraphics{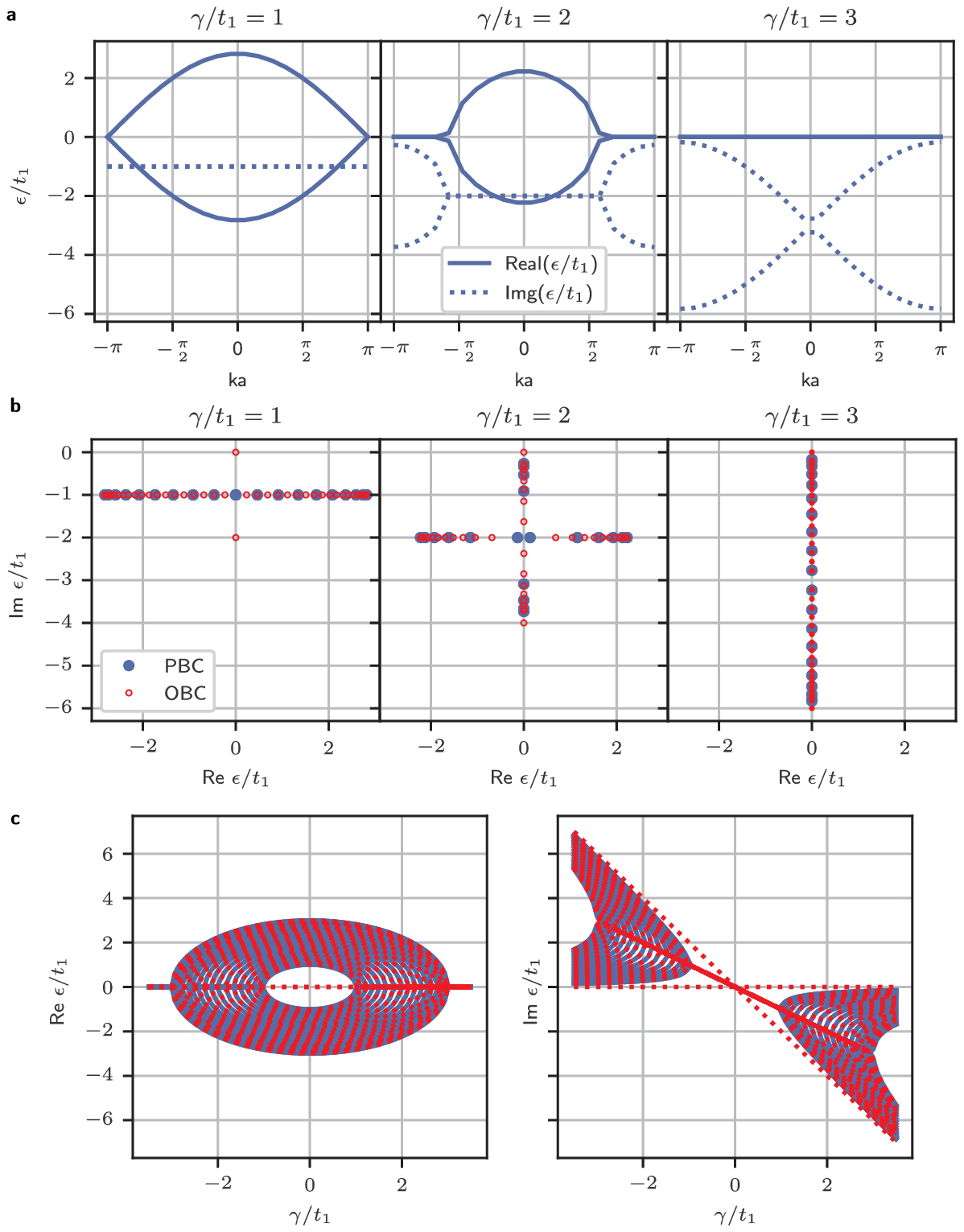}
    \caption{Spectrum of the SSH model with alternating loss for PBC and OBC in the topological regime $t_2 = 2 t_1$. A system size of 20 unit cells was used. \textbf{a}, As a function of $ka$ for different values of $\gamma$. \textbf{b}, In the complex plane for different values of $\gamma$. The blue full circles results correspond to PBC, and the red empty circles to OBC. \textbf{c}, As a function of $\gamma/t_1$. }
    \label{fig:l-topo}
\end{figure*}

Up until this point, we have been considering systems with gain as well as loss. In an experimental setting, gain would be more difficult to engineer than loss. In order to adapt the theory to this, we can transform the system of Eq.~\eqref{eq:hammatgainloss} to a system with only loss, by simply subtracting $i \gamma \mathbbm{1}$ from it. This results in
\begin{equation}
	 H =  \mathbf{c}^\dagger\begin{pmatrix}
 	-2\gamma i& -t_1 & 0 & \cdots & b\\ 
 	-t_1 & 0 & -t_2 &  & \\ 
 	0 & -t_2 & -2\gamma i &  & \vdots \\ 
 	\vdots&  &  & \ddots  & \\ 
 	 b&  & \cdots &  & 0
	\end{pmatrix}\mathbf{c},
	\label{eq:hammatloss}
\end{equation}
where $b$ is $-t_2$ (0) for PBC (OBC). This results in a spectrum very similar to the spectrum of Eq.~\eqref{eq:spectrumgainloss}, except that it has an imaginary offset
\begin{equation}
	\epsilon(k) = \pm\sqrt{-\gamma^2 + t_1^2 + t_2^2 + 2 t_1 t_2 \cos(ka)} - i \gamma.
	\label{eq:spectrumloss}
\end{equation}
This system can once again be studied in the trivial and topological regime.

\subsubsection{Trivial Phase}
This system is clearly very similar to the system with alternating gain and loss, as can also be seen from Figs.~\ref{fig:l-trivial}a-c. Figs.~\ref{fig:l-trivial}a-b are just shifted down on the imaginary axis proportional to $\gamma$, while Fig.~\ref{fig:l-trivial}c shows how the spectrum depends on $\gamma$. Due to this imaginary offset, the system no longer has a $\mathcal{PT}$-symmetric phase, but rather presents passive $\mathcal{PT}$-symmetry. Similarly to the system with alternating gain and loss, the behaviour of the spectrum does not change much for different boundary conditions.

\subsubsection{Topological Phase}
Like in the trivial case, the spectrum is very similar to the alternating gain and loss Hamiltonian, but just shifted down on the imaginary axis, as can be seen in Figs.~\ref{fig:l-topo}a-c. Again, there is no $\mathcal{PT}$-symmetric region due to the imaginary offset. The OBC case is discussed in detail in the main text.

\begin{figure*}[t]
    \centering
    \includegraphics{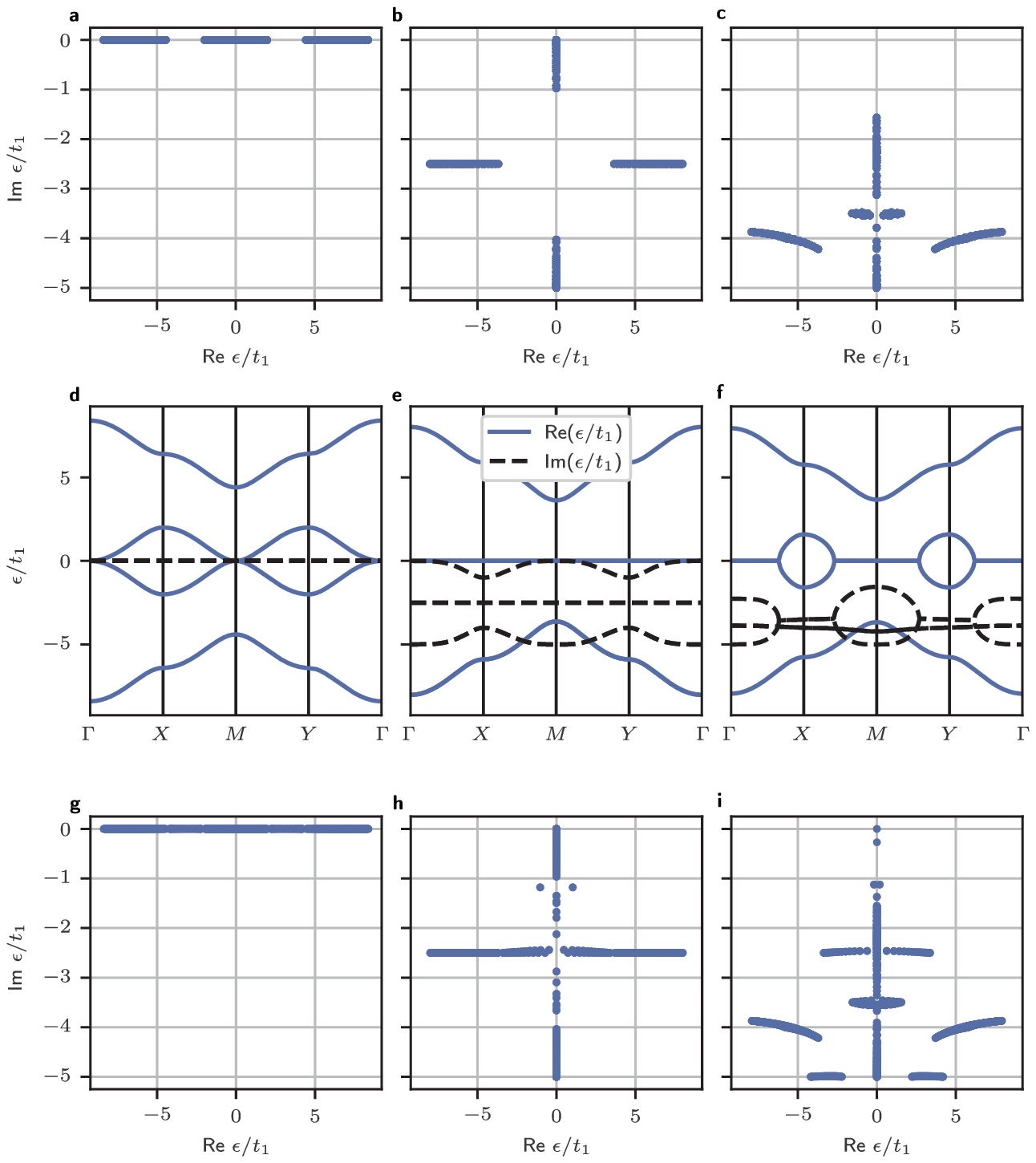}
    \caption{Spectrum of the 2D SSH model, for $t_2/t_1 = 3.2$. \textbf{a - c}, The spectrum in  the complex plane for PBC. \textbf{d - f}, Spectrum along the points of high symmetry for PBC. \textbf{g - i}, The spectrum in  the complex plane for OBC. \textbf{a}, \textbf{d}, and \textbf{g} show the spectrum for the lossless 2D SSH model. \textbf{b}, \textbf{e}, and \textbf{h} show the spectrum for the system with loss along the diagonal of the unitcell (with $\gamma/t_1 = 5$), yielding a bimode: only two modes remain at zero energy. \textbf{c}, \textbf{f}, and \textbf{i} show the spectrum for the model with loss at three sites in the unit cell (with $\gamma/t_1 = 5$), yielding a monomode, as there is only one mode remaining at zero energy.}
    \label{fig:s10}
\end{figure*}

\subsection{Alternating Loss 2D SSH}
The SSH model can also be extended into two dimensions. This model has the same basic behaviour as the 1D SSH model, exhibiting three phases: $\mathcal{PT}$-symmetric phase, $\mathcal{PT}$-broken phase and a phase with an imaginary line gap. In Fig.~\ref{fig:s10} we show the spectrum of the 2D SSH model with no loss (Fig.~\ref{fig:s10}a, d, and g) and with the loss distribution yielding a bimode (Fig.~\ref{fig:s10}b, e, and h) and a monomode (Fig.~\ref{fig:s10}c, f, and i). Especially Fig.~\ref{fig:s10}c makes it very clear that all the bulk modes are destroyed in this loss configuration. For OBC, this would leave only one edge state.

\section{Calculations regarding the Hermitian ancestor Hamiltonian}
\label{app:nonhem}
\subsection{Effective Hamiltonian for the 2D SSH}
\label{app:nonhem1}

The lossy 2D SSH model also has a Bloch Hamiltonian with non-Hermitian CS,  
\begin{equation}
    H_{\textrm{NH-2d-SSH}}(k_x, k_y)=\begin{pmatrix}
        i P&Q(k_x, k_y)\\
        Q^\dagger(k_x, k_y)&i R
    \end{pmatrix}
\end{equation}
with $Q(k)=\left[t_1+t_2\cos(k_x)\right]\mathbb{I}+i t_2\sin(k_x)\sigma_z+\left[t_1+t_2\cos(k_y)\right]\sigma_x- t_2\sin(k_y)\sigma_y$ using the choice of basis from Ref.~\cite{Benalcazar2020}. $R=-\gamma \mathbb{I}$, $P=0$ and $R=-\gamma \mathbb{I}$, $P=-\gamma\left(\mathbb{I}+\sigma_z\right)/2$ for the loss configuration of Fig.~4b and Fig.~4c, respectively, of the main text. 

We see that we can extend the method of Ref.~\cite{Brzezicki2019} to define an effective Hermitian Hamiltonian by adding a third dimension to the problem
\begin{equation}
\begin{split}
     H^\text{eff}(k_x, k_y,\eta) =&\ \Gamma \big[\eta \mathbb{I} - i H_{\textrm{NH-2D-SSH}}(k_x, k_y)\big]\\
     =&\ H_{d}(\eta)+[t_1+t_2\cos(k_x)]\sigma_y\otimes \mathbb{I}\\
     &+t_2\sin(k_x)\sigma_x\otimes \sigma_z\\
     &+[t_1+t_2\cos(k_y)]\sigma_y\otimes \sigma_x\\
     &-t_2\sin(k_y)\sigma_y\otimes \sigma_y,
\end{split}
\end{equation}
where $\Gamma=\sigma_z\otimes \mathbb{I}$ and the diagonal $H_d$ part of the Hamiltonian is given by, $H_d=-\gamma/2\,\mathbb{I}\otimes \mathbb{I}+(\eta-\gamma/2)\,\sigma_z\otimes \mathbb{I}$ and $H_d=-3\gamma/4\,\mathbb{I}\otimes \mathbb{I}+(\eta+\gamma/4)\,\sigma_z\otimes \mathbb{I}-\gamma/4\, \mathbb{I}\otimes \sigma_z -\gamma/4\, \sigma_z\otimes \sigma_z$ for the loss configuration in Fig.~4b and Fig.~4c of the main text, respectively. We show the dispersion relation of both cases in Fig.~5 of the main text. The corner modes of the 2D non-Hermitian model appear as linearly dispersing modes for the effective Hamiltonian, although they are not the only ones with this property because the bulk of this system is not gapped.

\subsection{Corner States of the Lossy 2D SSH Model}
\label{app:nonhem2}
The Jackiw-Rebbi construction can be generalized to describe corner states in higher-order topological phases \cite{Benalcazar2017, Ghosh2021} by considering a corner as an $x$-domain wall in the $y$-edge and vice-versa, and ensuring the compatibility of the two solutions. To use the continuum Hamiltonian with a position-dependent mass term, we consider that $t_1$ can be different along $x$ and $y$. We see then that the gap closing is given by $H_D=0$, $t_{1}^x=t_1^y=t_2$, and $k_x=k_y=\pi$. Therefore, we again obtain a low-energy continuum theory [Eq.~(27) of the main text]
\begin{widetext}
\begin{eqnarray}
     &&H^\text{eff}(p_x, p_y,\eta)=H_{d}(\eta)+t_2 M^x\sigma_y\otimes \mathbb{I}- t_2 p_x \sigma_x\otimes \sigma_z+t_2 M^y\sigma_y\otimes \sigma_x+t_2 p_y\sigma_y\otimes \sigma_y,
\end{eqnarray}
where $M^{x/y}\equiv \left(t_1^{x/y}-t_2\right)/t_2$. The higher-order topological phase is characterized by having both $M^x$ and $M^y$ negative, while the trivial phase has both positive. An edge of the HOT phase reproduces a domain wall between the HOT phase and the vacuum (which is trivial). To show that these states are topological, we focus on the upper right corner, which we locate at $x=y=0$, but the entire discussion is easily extended to the other three corners. 

A zero-energy mode in the right $x$-edge should satisfy 
\begin{equation}
    \left(t_2 M^x(x)\sigma_y\otimes\mathbb{I} +it_2\partial_x \sigma_x\otimes \sigma_z\right)\Psi^{x-\textrm{edge}}(x)=0=\left(M^x(x)\mathbb{I}\otimes\mathbb{I} +\partial_x \sigma_z\otimes \sigma_z\right)\Psi^{x-\textrm{edge}}(x).
\end{equation}
Writing $\Psi^{x-\textrm{edge}}(x)$ in terms of the eigenstates $\chi^{\sigma, \sigma'}$ of $\sigma_z\otimes \sigma_z$
\begin{eqnarray}
    &&\Psi^{x-\textrm{edge}}(x)=\sum\limits_{\sigma, \sigma'=\pm 1} c_{\sigma, \sigma'} \psi^{\sigma, \sigma'}_{x-\textrm{edge}}(x)\chi^{\sigma, \sigma'}, \quad \sum\limits_{\sigma, \sigma'}\left|c_{\sigma, \sigma'}\right|^2=1, \quad \int\limits_{-\infty}^{\infty} dx \left|\psi_{x-\textrm{edge}}^\sigma(x)\right|^2=1,\\
    &&\chi^{+, +}=\begin{pmatrix}
        1\\0\\0\\0
    \end{pmatrix},\quad \chi^{+, -}=\begin{pmatrix}
        0\\1\\0\\0
    \end{pmatrix},\quad \chi^{-, +}=\begin{pmatrix}
        0\\0\\1\\0
    \end{pmatrix}, \quad\chi^{-, -}=\begin{pmatrix}
        0\\0\\0\\1
    \end{pmatrix},
\end{eqnarray}
and using that the eigenvalues of $\sigma_z\otimes \sigma_z$ are $\sigma \sigma'$, we obtain the equation for $\psi_{x-\textrm{edge}}$
\begin{equation}
    \left(M^x(x)+\sigma\sigma'\partial_x \right)\psi^{x-\textrm{edge}}(x)=0\Rightarrow \psi^{x-\textrm{edge}}(x)=\mathcal{N}^x e^{-\sigma \sigma' \int\limits_{-\infty}^{x} M^x(x') dx'}.
\end{equation}
\end{widetext}
Setting the edge at $x=0$, we have a mass profile that has a negative (positive) value for negative (positive) $x$. Therefore, the solutions localized in the edge are given by $\sigma\sigma'=1$, which is satisfied by $\sigma=\sigma'=\pm 1$. For convenience, we use the subscript $+$ ($-$) in the first (second) solution. The components of the $x$-surface Hamiltonian are given by 
\begin{equation}
\begin{split}
    [H^{x-\text{surface}}(p_y)]_{l, m}=&\chi^{l \dagger}\left[ t_2 M^y\sigma_y\otimes \sigma_x+t_2 p_y\sigma_y\otimes \sigma_y\right]\chi^{m}\\
    &=t_2 M^y(\sigma_y)_{lm}+t_2 p_y (\sigma_x)_{lm} .
\end{split}
\end{equation}
Considering now the corner as being a ($y$) domain wall for this surface Hamiltonian, we look for a solution $\Psi^{\textrm{corner}}(y)$ which satisfies
\begin{equation}
    \left(t_2 M^y\sigma_y-i t_2 \partial_y \sigma_x\right)\Psi^{\textrm{corner}}(y)=0=\left(M^y\mathbb{I}- \partial_y \sigma_z\right)\Psi^{\textrm{corner}}(y).
\end{equation}
In terms of the spinors of the first solutions 
\begin{equation}
    \Psi^{\textrm{corner}}(y)=\sum\limits_{\tau=\pm 1} c_\tau \psi_{\textrm{corner}}^\tau(y) \chi^{\tau},
\end{equation}
we obtain 
\begin{widetext}
\begin{equation}
    \left(M^y(y)- \tau \partial_y \right)\psi_{\textrm{corner}}^\tau(y)=0\rightarrow \psi_{\textrm{corner}}^\tau(y)=\mathcal{N}^y e^{\tau \int\limits_{-\infty}^{y} M^y(y') dy'}.
\end{equation}
Therefore, we see that the localized solution is the one with $\tau=-1$, which corresponds to $\chi^{-}$. Accordingly, the wavefunction for the corner state is given by
\begin{equation}
    \Psi^{\textrm{corner}}(x, y, \eta)=\mathcal{N}^x\mathcal{N}^y e^{- \int\limits_{-\infty}^{x} M^x(x') dx'}e^{-\int\limits_{-\infty}^{y} M^y(y') dy'}\psi(\eta) \begin{pmatrix}
        0\\0\\0\\1
    \end{pmatrix}. 
\end{equation}

We have then the corner Hamiltonian
\begin{equation}
    H^{\textrm{corner}}(\eta)=\langle \Psi^{\textrm{corner}}|H^{\text{eff}}|\Psi^{\textrm{corner}}\rangle=\chi^{+\dagger}H_d(\eta)\chi^+=-\eta+\gamma
\end{equation}
\end{widetext}
for both loss configurations. The same construction can be used to obtain the other corner modes, the only difference being how $M^{x/y}$ changes across the boundaries, which affects both the spatial and spinorial forms of the corner states. 

\section{Numerical Predictions}
\label{app:numpred}
 \begin{figure*}[t]
    \centering
    \includegraphics[]{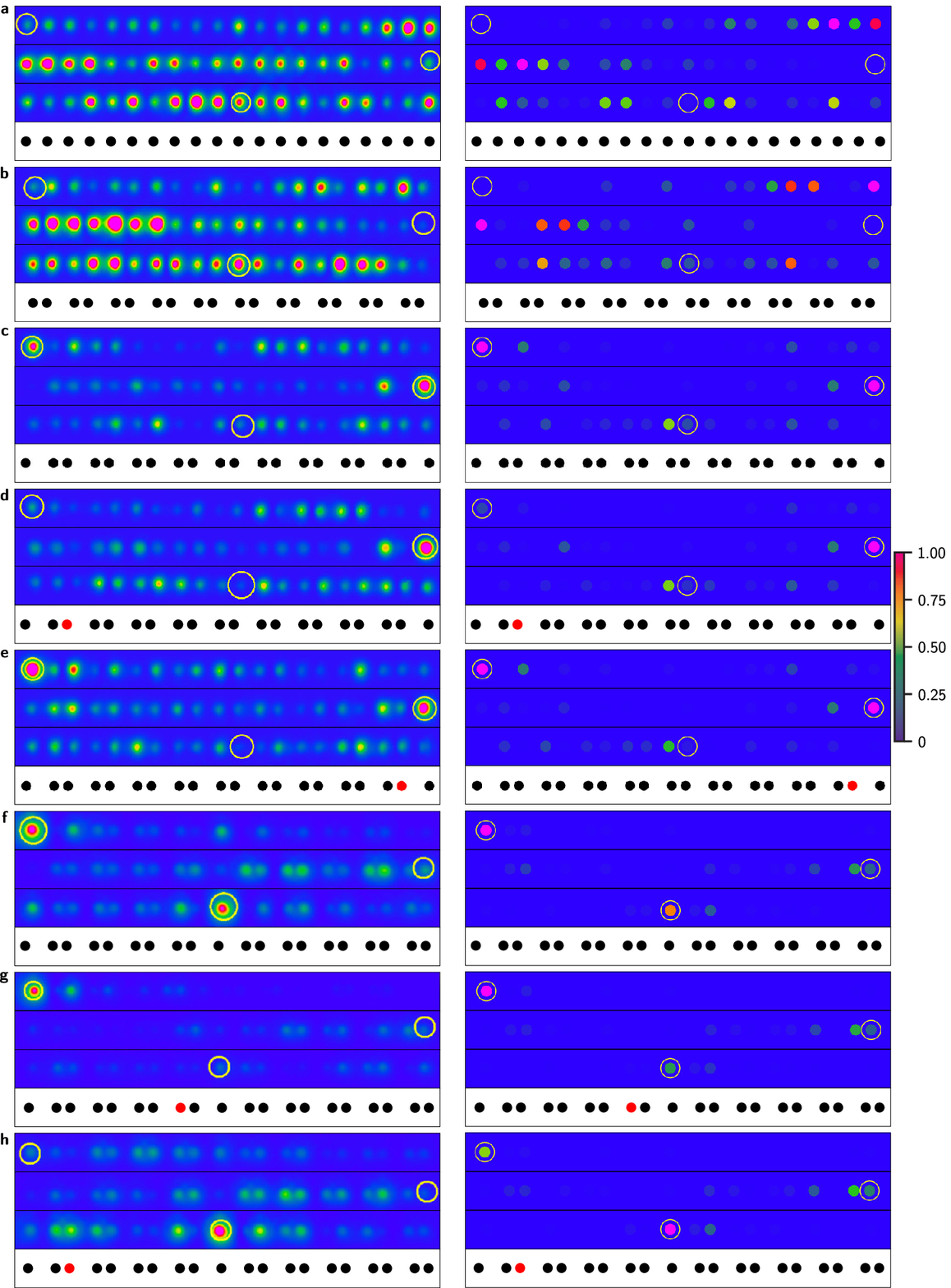}
    \caption{Numerical results of Fig.~3 of the main text, alongside the experimental data. The left column gives the experimental data, which can also be seen in the main text. The right column shows the numerical results. Yellow circles signal the point at which light was injected. \textbf{a}, Tight-binding chain (black dots) with $t_1 = t_2$ dispersing into the bulk. \textbf{b}, Trivial SSH chain with $t_1/t_2 = 2$ dispersing into the bulk. \textbf{c}, Topological SSH chain without loss with $t_2/t_1 =2$. Localised edge modes can be clearly observed. \textbf{d}, When we add loss ($\gamma/t_1 = 0.5$) at the red-dot site near the left edge, one of the edge modes disappears, revealing the monomode. \textbf{e}, This is not the case when the loss is applied far away from the left edge. In this case, the edge mode does not decay within the experimental time scale. \textbf{f}, A topological defect is added to the system ($t_2/t_1 =3.2$). Now, there is a left edge mode, as well as a mode pinned on the defect. \textbf{g}, Loss is placed on the red-dot site near the defect. This destroys the mode at the defect, but leaves the monomode. \textbf{h}, By placing the loss on the red-dot site near the left edge, the edge mode gets destroyed, but not the defect mode.}
    \label{fig:S6}
 \end{figure*}
\begin{figure*}[t]
    \centering
    \includegraphics[]{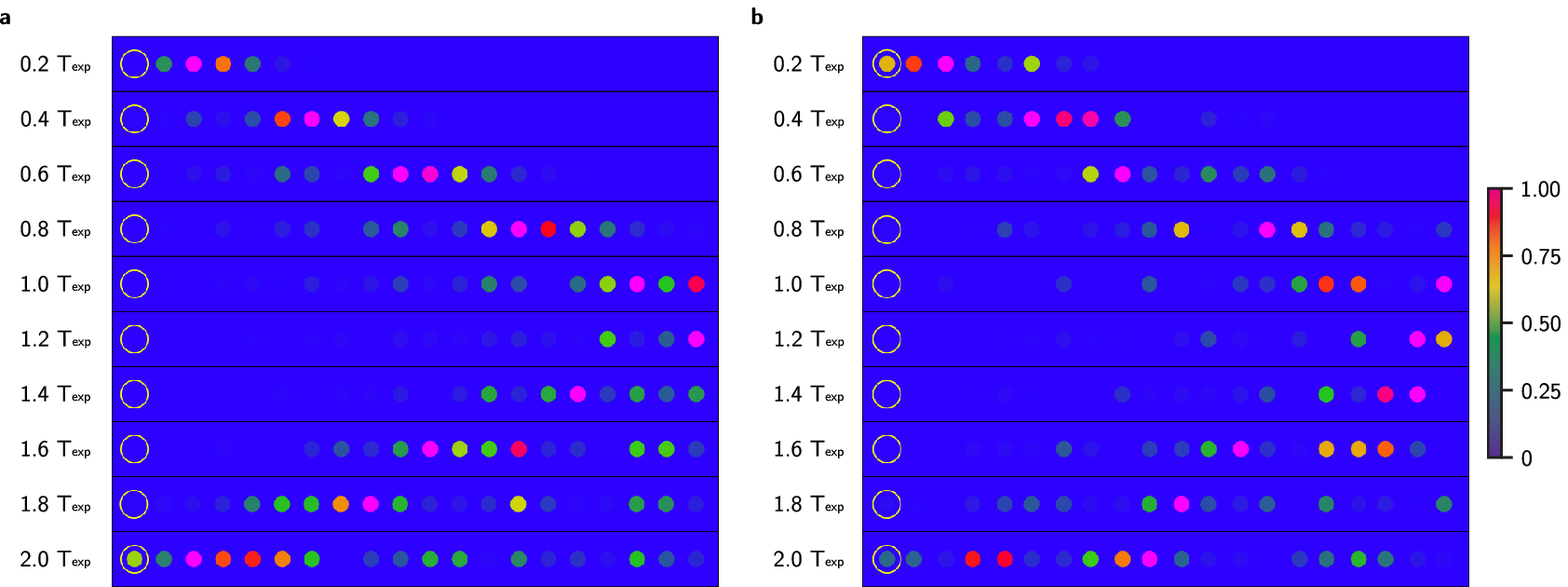}
    \caption{Numerical predictions of the time propagation of Figs.~3a-b. \textbf{a}, Time propagation of the tight-binding chain. Clearly, a wave is propagating from its insertion point on the left, to the right. At around the time of the experiments, the wave hits the right edge of the chain and reflects back. \textbf{b}, Time propagation of the trivial SSH chain. We see a similar behaviour as for \textbf{a}.}
    \label{fig:S7}
\end{figure*}
\begin{figure*}[t]
    \centering
    \includegraphics[]{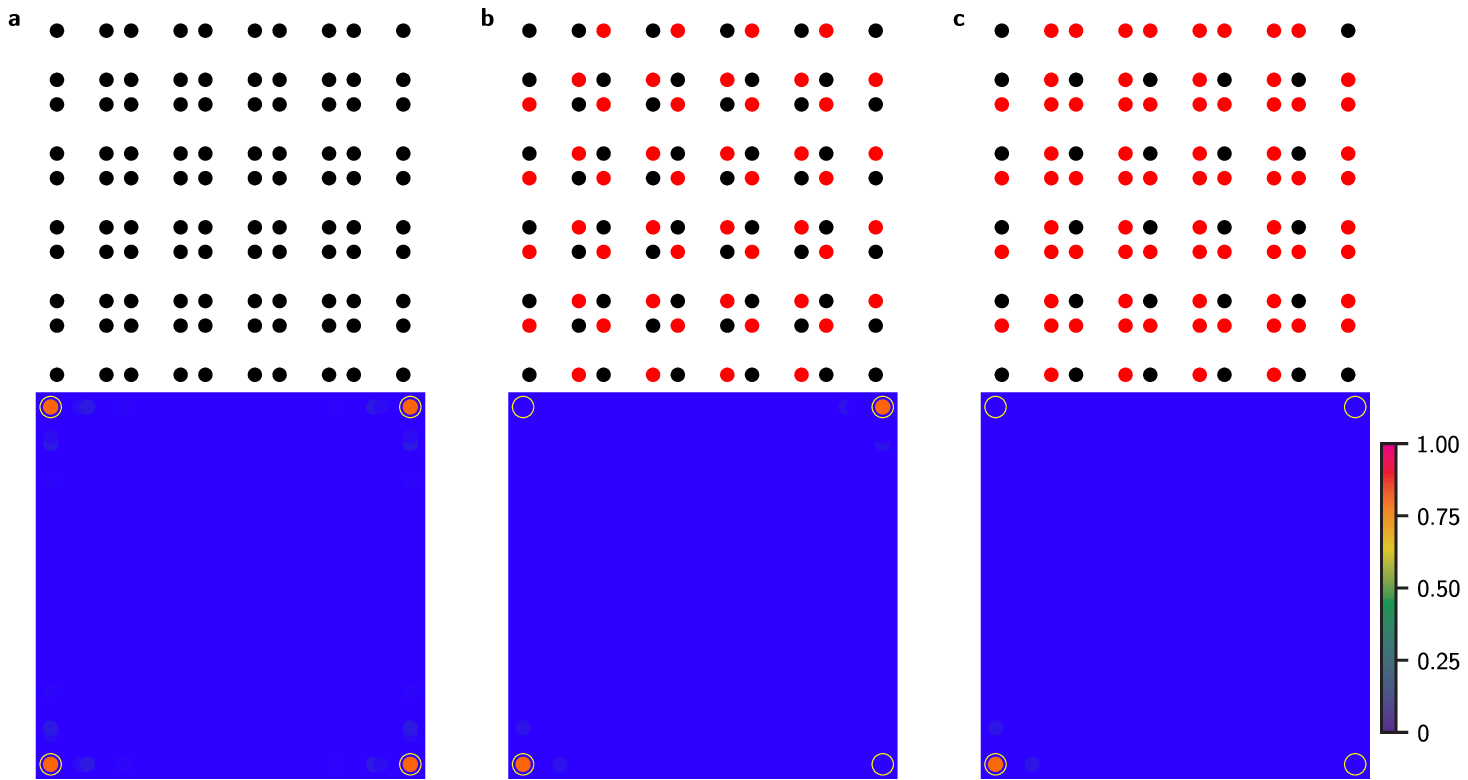}
    \caption{Numerical prediction of the bi- and monomodes in the 2D SSH model for 10x10 unit cells with $t_2/t_1 = 3.2$ and $\gamma/t_1 = 5$. Yellow circles indicate the point at which light was injected. The top row shows the loss locations in the lattice (red dots) and the bottom row the predictions, for injecting into the four corners. \textbf{a}, No loss, showing four distinct corner modes. \textbf{b}, Loss on two sublattices. This breaks the $C_{4v}$ symmetry of the system into $C_{2v}$ and leads to bimodes. \textbf{c}, By adding loss on three sublattices, the $C_{2v}$ symmetry is broken and the monomode emerges.}
    \label{fig:S8}
\end{figure*}

Fig.~\ref{fig:S6} shows the experimental results (first column) alongside the theoretical predictions for the 1D SSH experiments (second column). These are the same numerical results as represented in Fig.~3 by circles, but here they are shown on a similar color scale to the experiments. Fig.~\ref{fig:S7} displays the time-evolution of the inserted light for the simple tight-binding case (Fig.~\ref{fig:S7}a), and the trivial SSH case (Fig.~\ref{fig:S7}b). Due to the paraxial equation, we can relate the length of the waveguides to time. As light moves along the waveguide, time can be promptly determined by the length divided by the speed of light. Therefore, this spatial direction actually corresponds to a time direction. The $T_{exp}$ can be calculated by scaling the length of the waveguide by the hopping parameter $t_1$, yielding $T_{exp} = L t_1$. This makes it clear that the theoretical predictions for these two models rely heavily on the duration of the experiment (length of the waveguide). Fig.~\ref{fig:S8} shows the theoretical results for the 2D SSH model. Here, a larger system size was used than in the experiment. However, the same behaviour can be seen, with the realization of a bimode and a monomode.

\section{Additional Experimetal Data}
\begin{figure*}[t]
    \centering
    \includegraphics{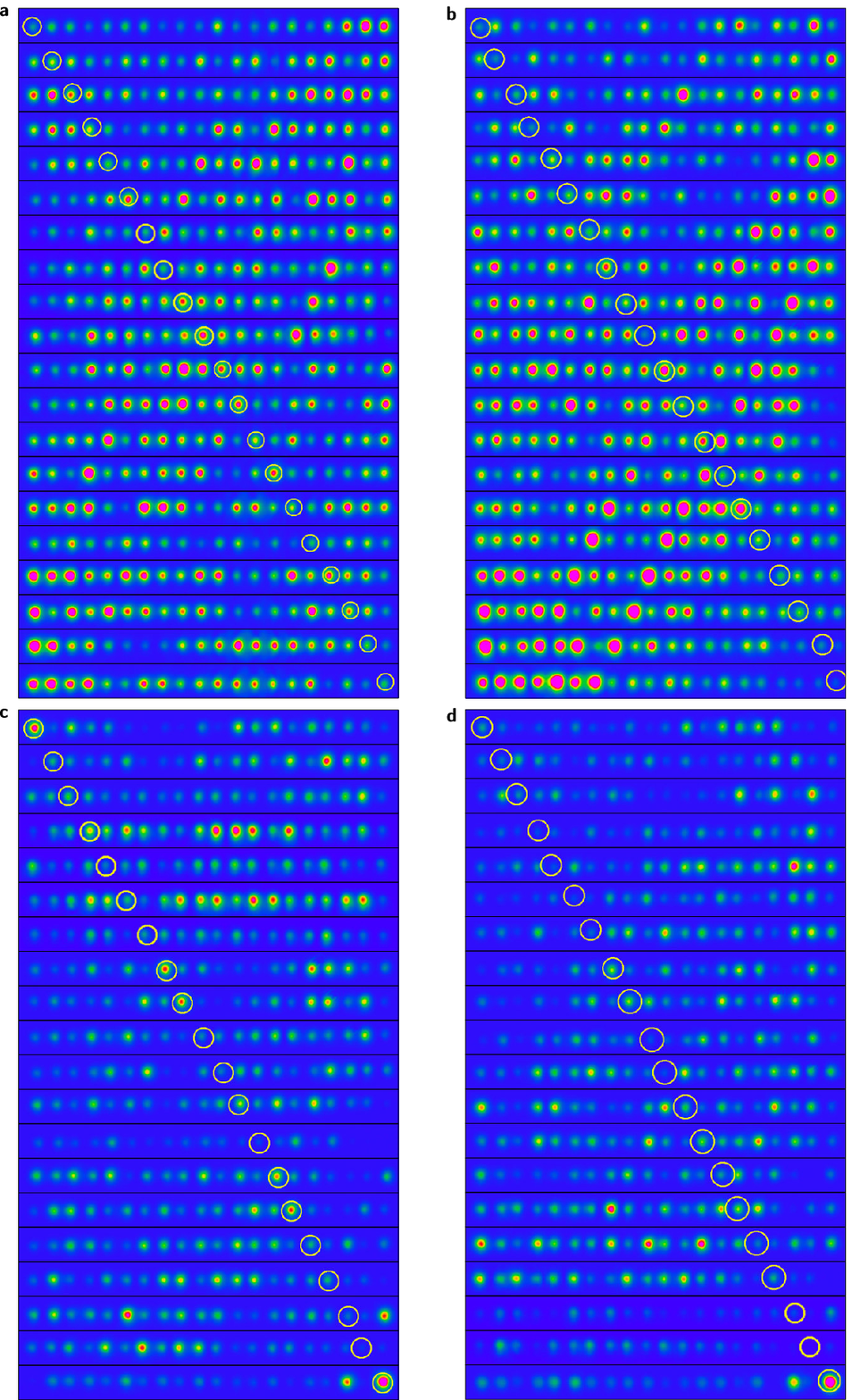}
    \caption{Experimental results for all insertion points. \textbf{a}, Results of the tight-binding chain with no loss, from Fig.~3a. \textbf{b}, Results of the topologically trivial chain with no loss, from Fig.~3b. \textbf{c}, Results of the topological chain with no loss, from Fig.~3c. \textbf{d}, Results of the topological chain with loss, from Fig.~3d.}
    \label{fig:S11}
\end{figure*}

\begin{figure*}[t]
    \centering
    \includegraphics{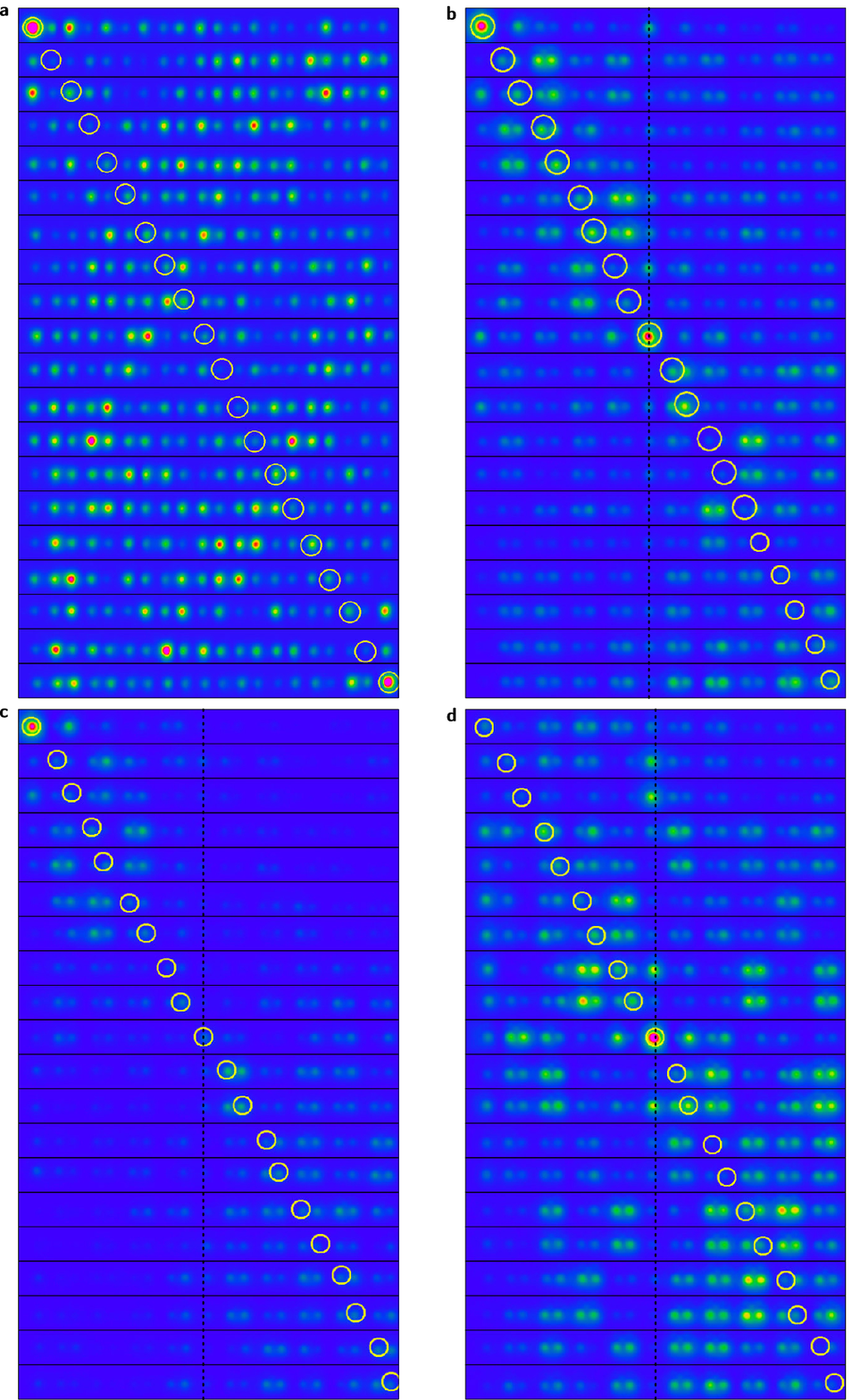}
    \caption{Experimental results for all insertion points. \textbf{a}, Results of the topological chain with loss on the "wrong" (far away) side of the lattice, from Fig.~3e. \textbf{b}, Results of the topological defect chain with no loss, from Fig.~3f. The position of the topological defect is indicated by a dotted line. \textbf{c}, Results of the topological defect chain with loss engineered in a way to destroy the defect mode, from Fig.~3g. \textbf{d}, Results of the topological defect chain with loss designed to destroy the left edge mode, from Fig.~3h.}
    \label{fig:S9}
\end{figure*}

Figs.~\ref{fig:S11} and \ref{fig:S9} show the full experimental data of the 1D chain experiments, obtained upon inserting light in each waveguide.

\clearpage

\bibliography{mainnew}

\end{document}